\newif\ifShowKeys
\ShowKeystrue
\ShowKeysfalse

\documentclass[11pt]{article}
	\pdfoutput=1
\usepackage[usenames,dvipsnames]{xcolor}
\usepackage[setpagesize=false,pagebackref=false, linktocpage, bookmarksopen=true, colorlinks=true, linkcolor=Maroon,citecolor=Maroon,urlcolor=Maroon]{hyperref}
\usepackage[parsep]{collref}

\ifShowKeys \usepackage[notcite]{showkeys} \fi

\usepackage{amsmath, amssymb,amsthm}
\numberwithin{equation}{section}

\usepackage{bm,environ,mathrsfs,array,arydshln}
\usepackage{booktabs,float,slashed,hyperref}
\usepackage[mathcal]{euscript}
\usepackage{tensor} 	
\usepackage{mathabx}
\usepackage[vcentermath]{youngtab}
\usepackage{xcolor}

\usepackage{aurical}
\usepackage[T1]{fontenc}
\usepackage[nodayofweek]{date time}
\usepackage{graphicx,epsfig,epic}
\usepackage{tikz} 
\usetikzlibrary{arrows,decorations.pathreplacing,decorations.markings,snakes}
\usetikzlibrary{cd}

\usepackage{framed}						
\definecolor{shadecolor}{rgb}{0.9996078, 0.984314, 0.960784}

\usepackage{comment}

\allowdisplaybreaks

\definecolor{myred}{RGB}{233, 33, 45}

\newcommand{\bs}{\begin{shaded}}
\newcommand{\es}{\end{shaded}\noindent}
\def\ba#1\ea{\begin{align}#1\end{align}}		
\newcommand{\be}{\begin{equation}}
\newcommand{\ee}{\end{equation}}
\newcommand{\bea}{\begin{equation} \begin{aligned}} 
\newcommand{\eea}{\end{aligned} \end{equation}}
\newcommand{\mc}{\mathcal }

\newcommand{\la}{\label}

\newcommand{\lp}{\notag \\ & }

\newcommand{\cf}{\textit{cf.} }
\newcommand{\ie}{\textit{i.e.} }
\newcommand{\eg}{\textit{e.g.} }

\newcommand{\N}{\mathcal N}

\newcommand{\mg}{m_{\Gamma}}

\DeclareMathOperator{\Tr}{Tr}

\newcommand{\mk}{\mathfrak}

\newcommand{\I}{\mathrm{I}}
\newcommand{\Isp}{\widehat{\I}}

\newcommand{\E}{\mathbb{E}}

\newcommand{\wei}[3]{P_{#1}\begin{bmatrix} #2 \\ #3\end{bmatrix}}
\newcommand{\vth}{\vartheta}

\begin{document}




\begin{titlepage}

\vspace*{15mm}
\begin{center}
{\Large\sc   Giant graviton expansion of Schur index}\vskip 9pt
{\Large\sc      and quasimodular forms}

\vspace*{10mm}

{\Large M. Beccaria and A. Cabo-Bizet}

\vspace*{4mm}
	
Universit\`a del Salento, Dipartimento di Matematica e Fisica \textit{Ennio De Giorgi},\\ 
		and I.N.F.N. - sezione di Lecce, Via Arnesano, I-73100 Lecce, Italy
			\vskip 0.3cm
\vskip 0.2cm {\small E-mail: \texttt{matteo.beccaria@le.infn.it, acbizet@gmail.com}}
\vspace*{0.8cm}
\end{center}

\begin{abstract}  
The flavored superconformal Schur  index of $\N=4$ $U(N)$ SYM has finite $N$ corrections encoded in its giant graviton expansion 
in terms of D3 branes wrapped in $AdS_{5}\times S^{5}$. The key element of this 
decomposition is the non-trivial index of the theory living on the wrapped brane system. A remarkable feature of the Schur limit
is that the brane index is an analytic continuation of the flavored index of $\N=4$ $U(n)$ SYM,  
where $n$ is the total brane wrapping number. We exploit recent exact results about the Schur  index of $\N=4$ $U(N)$ SYM 
to evaluate the closed form of the the brane indices appearing in the giant graviton expansion. 
Away from the unflavored limit, they are characterized by quasimodular forms providing exact information at all orders in the index universal fugacity. 
As an application of these results, we 
present novel exact expressions for the giant graviton expansion of the unflavored Schur index in a class of four dimensional $\N=2$ theories
with equal central charges $a=c$, \ie the  non-Lagrangian theories $\widehat{\Gamma}(SU(N))$ with $\Gamma = E_{6}, E_{7}, E_{8}$.
\end{abstract}
\vskip 0.5cm
	{
	}
\end{titlepage}

\tableofcontents
\vspace{1cm}


%

\section{Introduction}

For a superconformal theory the index introduced in \cite{Kinney:2005ej,Romelsberger:2005eg,Romelsberger:2007ec} 
encodes a large amount of information about the BPS spectrum. It may be regarded as the Witten index \cite{Witten:1982df} in radial quantization and is invariant 
under smooth supersymmetric deformations. Indices are nowadays a standard and major tool for the investigation of superconformal theories \cite{Argyres:2022mnu}. Their applications range from the study of their dualities, strongly coupled fixed points, and even black hole physics in the context of AdS/CFT, see \eg in $AdS_4$ \cite{Benini:2015eyy, choi2019quantum, Nian:2019pxj} and in $AdS_5$~\cite{Cabo-Bizet:2018ehj,Choi:2018hmj,Benini:2018ywd}. A review can be found in \cite{Zaffaroni:2019dhb}.

For $U(N)$ gauge invariant theories, the index admits a large $N$ limit with finite $N$ corrections 
coming from $U(N)$ trace relations that spoil orthogonality of single-trace operators. When the superconformal theory has a gravity dual,
the interpretation of these corrections is far deeper. In the simplest example of four dimensional $\N=4$ $U(N)$ super Yang-Mills (SYM) theory dual to type IIB superstring in $AdS_{5}\times S^{5}$, 
finite $N$ effects originate from wrapped D3 brane states maximally stretched in $S^{5}$ called giant gravitons \cite{McGreevy:2000cw} and having charge of order $N$. \footnote{
See \cite{Lee:2023iil} 
for a recent analysis of the precise correspondence between states of certain D3 giant gravitons branes in $AdS_{5}\times S^{5}$ and auxiliary ghost states in $\N=4$ $U(N)$ SYM.}
From the corresponding giant-graviton expansion of the index, it is possible to reproduce the entropy of dual black holes in $AdS_{5}\times S^{5}$ as first shown in \cite{Choi:2022ovw} for small black holes with charge $Q\ll N^{2}$ and later generalized to  black holes
with charge~$Q\sim N^{2}$, first in \cite{Beccaria:2023hip}, 
using a large-charge expansion, and later in~\cite{Kim:2024ucf}, using a large $N$ expansion. Recently, the giant graviton expansion has also been described by enumerating BPS geometries
studying bubbling solutions in supergravity  \cite{Chang:2024zqi, DeddoLiu}.

The contribution to the index from a certain wrapped D3 brane state may be identified with  the  index of the theory living on the brane world-volume and having reduced unbroken superconformal
invariance \cite{Imamura:2021ytr,Gaiotto:2021xce,Lee:2022vig}. The computation of this ``brane index'' is  a non-trivial  issue.
The superconformal index of 4d  $\N=4$ $U(N)$ SYM is defined as 
\be
\la{1.1}
\I^{U(N)}(y, \bm{u}; q) = \Tr_{\rm BPS}[(-1)^{\rm F}\, q^{H+\bar J}y^{2J}u_{1}^{R_{1}}u_{2}^{R_{2}}u_{3}^{R_{3}}], \qquad u_{1}u_{2}u_{3}=1\, ,
\ee
where the Hamiltonian $H$, the spins $J, \bar J$, and the R-charges $R_{1}, R_{2}, R_{3}$ are Cartan generators of the superconformal algebra $\mk{psu}(2,2|4)$
and the trace is restricted to  states obeying certain BPS conditions.
\footnote{We follow the conventions in 
 \cite{Arai:2019xmp} where, in particular,  details on the BPS condition can be found.}  The giant graviton expansion of the index reads \cite{Imamura:2021ytr,Gaiotto:2021xce}
\be
\la{1.2}
\I^{U(N)}(y, \bm{u}; q)  = \I^{\rm KK}(y, \bm{u}; q)\, \sum_{n_{1}, n_{2}, n_{3}=0}^{\infty} (qu_{1})^{n_{1}N}(qu_{2})^{n_{2}N}(qu_{3})^{n_{3}N}\, \I_{n_{1},n_{2},n_{3}}^{\rm D3}(y, \bm{u}; q)\, ,
\ee
where each piece has a definite meaning on gravity dual side. The $N\to\infty$ factor $\I^{\rm KK}(y, \bm{u}; q)$ matches the  index from IIB supergravity Kaluza-Klein  states. The triple summation is over wrapped D3 branes with topology  $S^{1}\times S^{3}$
where $S^{1}$ is in $AdS_{5}$ and $S^{3}\subset S^{5}$ with winding $n_{I}$  along the 3-cycle obtained setting $\mathsf{z}_{I}=0$ in $S^{5}$ 
realized  in $\mathbb{C}^{3}$ as  $|\mathsf{z}_{1}|^{2}+|\mathsf{z}_{2}|^{2}+|\mathsf{z}_{3}|^{2}=1$.
The  quantities $\I_{n_{1},n_{2},n_{3}}^{\rm D3}(y, \bm{u}; q)$ represent  the superconformal index of the theory living on the wrapped branes world-volume, \ie 
a  $U(n_{1}) \times U(n_{2})\times  U(n_{3})$  quiver gauge theory with three bi-fundamental hyper multiplets. The prefactor $\prod_{I=1}^{3}(qu_{I})^{n_{I}N}$ in (\ref{1.2})
comes from the classical charges and energy of the  brane system.

At generic values of fugacities $y, \bm{u},q$, the brane indices $\I_{n_{1},n_{2},n_{3}}^{\rm D3}$ do not depend on $N$ and (\ref{1.2}) organizes finite $N$ corrections $\sim q^{nN}$
as a  series of  contributions from  wrapped branes  with increasing total winding number. \footnote{
It is possible to discuss (\ref{1.2}) in the $U(N)$ SYM theory without reference to its string dual description \cite{Murthy:2022ien,Liu:2022olj,Eniceicu:2023cxn}, up to possible non-trivial reorderings of the sum \cite{Eniceicu:2023uvd}.
} At large $N$, the corrections are non-perturbative in the small $q$ expansion. 
When algebraic constraints are imposed on fugacities to get simplified unrefined indices, extra polynomial contributions in $N$ may appear. This is often referred to as a ``wall-crossing'' effect.
It was discussed in \cite{Gaiotto:2021xce,Lee:2022vig,Beccaria:2023zjw} and  explained in \cite{Beccaria:2023cuo,Beccaria:2024vfx} on gravity side in terms of zero modes of fluctuations on the wrapped branes, see also \cite{Gautason:2023igo}.
These zero modes are fully  regularized when all  fugacities are switched on and take generic values.

In this paper, we focus on some technical aspects of the determination of the brane indices $\I_{n_{1},n_{2},n_{3}}^{\rm D3}$ in (\ref{1.2}).
To this aim,  it will be convenient to consider the so-called Schur limit of the index \cite{Gadde:2011ik,Gadde:2011uv}. 
For a general $\N = 2$ superconformal theory, the Schur index gives the vacuum character of the associated chiral algebra \cite{Beem:2013sza}. Here, 
it corresponds to  the specialization $y=q^{\frac{1}{2}}$, $\bm{u} = (u,u^{-1},1)$ of (\ref{1.1}), \ie  \footnote{To keep notation simple, the Schur index is denoted 
by the same symbol as the general index and only two arguments $u, q$ instead of the five $y, u_{1}, u_{2}, u_{3}, q$.}
\be
\la{1.3}
\I^{U(N)}(u; q) = \Tr_{\rm BPS}[(-1)^{\rm F}\, q^{H+J+\bar J}u^{R_{1}-R_{2}}]\, .
\ee
The specific structure of the giant-graviton expansion of the Schur index of $\N=4$ $U(N)$ SYM  was first discussed in \cite{Arai:2020qaj}.
Quite remarkably, in this limit the right hand side of (\ref{1.2}) can be written in terms of the SYM index itself, up to a tricky analytic continuation 
proposed in \cite{Arai:2019xmp}. The explicit formula reads 
\ba
\la{1.4}
\I^{U(N)}(u; q) = \I^{\rm KK}(u;q)\,  \sum_{n=0}^{\infty}\sum_{p=0}^{n} (uq)^{(n-p)N}\I^{\rm D3}_{n-p}(u; q) q^{2(n-p)p}(u^{-1}q)^{pN}\I^{\rm D3}_{p}(u^{-1}; q)\, ,
\ea
where the  Kaluza-Klein factor is explicitly (see Appendix \ref{app:special} for special functions notation)
\be
\la{1.5}
 \I^{\rm KK}(u;q) = \frac{(q^{2})_{\infty}}{(uq)_{\infty}(u^{-1}q)_{\infty}}\, ,
\ee
and the brane indices are given by the key relation
\be
\la{1.6}
\I^{\rm D3}_{n}(u; q) = \I^{U(n)}(u^{-\frac{1}{2}}q^{-\frac{3}{2}}; u^{-\frac{1}{2}}q^{\frac{1}{2}})\,.
\ee
Thus, finite $N$ corrections $\sim q^{nN}$ to the Schur index of  $U(N)$ SYM  are captured by  the  index of the $U(n)$ SYM theory  by means of the above fugacity 
transformation. As a technical  remark, the reduction of the triple sum in (\ref{1.2}) to the single sum in (\ref{1.4}) is a simplification 
due to vanishing of contribution from the cycle $\mathsf{z}_{3}=0$ and a 
Weyl symmetry $u\to u^{-1}$ relating contributions from the other two cycles $\mathsf{z}_{1}=0$ and $\mathsf{z}_{2}=0$.

The analytic continuation rule  (\ref{1.6}) follows from the analysis of the unbroken superconformal symmetry on the wrapped brane system. 
This symmetry-based approach to the construction of brane indices 
was recently reconsidered in \cite{Beccaria:2024vfx,Gautason:2024nru} by a full calculation of fluctuation effects from all fields living on the brane world-volume. 
A similar treatment of finite $N$ corrections to superconformal index was previously carried out in the case of the 6d $(2,0)$ theory from semiclassical M2 brane
wrapped on $S^{1}\times S^{2}$ in the M-theory background $AdS_{7}\times S^{4}$ \cite{Beccaria:2023sph}, and for the
3d $\N=8$ supersymmetric level-one $U(N)\times U(N)$ ABJM theory from semiclassical M5 brane wrapped on $S^{1}\times S^{5}$ in  $AdS_{4}\times S^{7}$  \cite{Beccaria:2023cuo}.
Alternatively, localization on the wrapped brane was recently used in \cite{Eleftheriou:2023jxr} to discuss the giant graviton expansion of the $\frac{1}{2}$-BPS conformal index in $\N=4$ $U(N)$ SYM.

Technically, a major problem in making use of   (\ref{1.6}) is that it cannot be applied when 
the $U(n)$ Schur index is only given as a (truncated) $q$-series. Reason is that the transformation (\ref{1.6}) is actually
an analytic continuation 
outside the convergence domain of the Schur index around $q=0$.
As a consequence, the strategy adopted in  \cite{Arai:2020qaj}  to overcome this difficulty was to \textit{(i)} apply the transformation
$(u; q)\to (u^{-\frac{1}{2}}q^{-\frac{3}{2}}; u^{-\frac{1}{2}}q^{\frac{1}{2}})$ to the integrand of the holonomy integral representation of the $U(n)$ index and \textit{(ii)} perform the
 contour integrations along suitable cycles by a prescription determining the set of relevant poles.  The same technique was successfully applied in a variety of other models in 
\cite{Arai:2019wgv,Arai:2019aou,Arai:2020uwd,Fujiwara:2021xgu,Imamura:2021dya,Fujiwara:2023bdc} and further developed and clarified in \cite{Imamura:2022aua,Fujiwara:2023bdc}.

The outcome of the symmetry-based approach were explicit expansions for the brane indices in (\ref{1.6}). 
At leading and next-to-leading wrapping order, \ie for $n=1,2$, they take the form 
\bea
\la{1.7}
\I^{\rm D3}_{1}(u;q) &= \frac{u^{3}}{1-u^{2}}\,q+(1-u^{2})q^{2}+(u^{-3}-u^{3})\, q^{3}+\cdots\, , \\
\I^{\rm D3}_{2}(u;q) &= \frac{u^{10}(2-u^{2})}{(1-u^{2})(1-u^{4})}\,q^{4}+u^{5}\,(2-u^{4})\,q^{5}+(2+2u^{6}-u^{12})\, q^{6}+\cdots\, ,
\eea
and were computed  in \cite{Arai:2020qaj} up to wrapping $n=4$. 
Expressions  (\ref{1.7}) have a very rich dependence on $u$ and $q$ that can be (minimally) tested by matching the exact prediction for the index in the unflavored case
$u=1$. This was obtained in  \cite{Bourdier:2015wda} and  reads
\bea
\la{1.8}
\I^{U(N)}(1; q) &= \I^{\rm KK}(1; q)\,\sum_{n=0}^{\infty}(-1)^{n}\bigg[\binom{N+n}{N}+\binom{N+n-1}{N}\bigg]\,q^{nN+n^{2}}\,, \\
&=  \I^{\rm KK}(1; q)\,\bigg[1-(N+2)\, q^{N+1}+\frac{1}{2}(1+N)(4+N)\, q^{2N+4}+\cdots\bigg]\, ,
\eea
where the Kaluza-Klein contribution is given by 
\be
 \I^{\rm KK}(1; q) = \frac{1}{\vartheta_{4}(0)} =  \frac{(q^{2})_{\infty}}{(q)_{\infty}^{2}} = \prod_{n=1}^{\infty}\frac{1+q^{n}}{1-q^{n}}\,.
\ee
Checking agreement between (\ref{1.4}) and (\ref{1.8}) turns out to be a non-trivial issue, because of the poles at $u=1$ in the $q$-series of the brane indices (\ref{1.7}). 
These have to cancel, but produce leftover powers of $N$, well visible in (\ref{1.8}),  due to the mentioned wall-crossing effect. 

Besides,  
higher order terms of the expansion do not follow from an explicit closed formula and are not under full analytical control.  Still, truncating the $q$-series at some high finite order, one 
empirically matches (\ref{1.8}) \cite{Arai:2020qaj}.

\paragraph{Summary of new results}

In this paper, we begin by considering the $u\to 1$ limit of (\ref{1.4}) at all orders in $q$. To this aim, we exploit recent exact results about the flavored  Schur index  
of $\N=4$ $U(N)$ SYM
\cite{Pan:2021mrw,Hatsuda:2022xdv}, see also \cite{Du:2023kfu} 
for other gauge groups. We show that its available explicit representations  allow to evaluate the analytic continuation in (\ref{1.6})
and to get closed form expressions for the brane indices $\I_{n}^{\rm D3}(u;q)$ in agreement with the expansions (\ref{1.7}), and extending them at all orders in $q$.  
As a consequence, we prove that the $u\to 1$ limit of (\ref{1.4})
reproduces the exact result (\ref{1.8}). In more details, we find that the general structure of the brane index is 
\ba
\I^{\rm D3}_{n}(u; q) &= \frac{u^{n(2n+1)}A_{n}(u)}{\prod_{m=1}^{n}(1-u^{2m})}q^{n^{2}}+\frac{B_{n}(u; q)}{(1-u^{2})^{n-2}}\,q^{n^{2}+1} \, ,
\ea
where $A_{n}(u)$ are  polynomials in $u$ with degree $n(n-1)$ and integer coefficients that we compute explicitly for all $n$. 
The functions $B_{n}(u; q)$ have  $q$-series with coefficients that are rational functions of $u$, smooth for $u\to 1$. 
They can be computed exactly near the point $u=1$ and have the structure
\be
\la{1.11}
B_{n}(u;q) = \frac{1}{q}\sum_{m=0}^{\infty}\frac{1}{m!}B_{n}^{(m)}(q)\, (u-1)^{m}\, ,
\ee
where $B_{n}^{(m)}(q)$ are quasimodular forms, \ie polynomials in the classical Eisenstein series $\E_{2}(q)$, $\E_{4}(q)$, $\E_{6}(q)$. Similar quasimodular properties 
were previously observed in the unflavored
Schur indices of class $\mc S$ theories \cite{Beem:2021zvt,Pan:2021mrw}, see also \cite{Razamat:2012uv,Beem:2017ooy,Huang:2022bry,Zheng:2022zkm}.

From these expressions, one can extend systematically the unrefined expansion  (\ref{1.8}) by including near-unflavored corrections in powers of the deviation $u-1$, and correct to all orders in $q$. 
To give an example, 
the leading term in the giant graviton expansion, \ie the correction with weight $\sim q^{N}$,  reads
\be
\I^{U(N)}(u; q) = \I^{\rm KK}(1; q)\,\bigg[1+q^{N+1}\bigg(-(2+N)+\sum_{p=2}^{\infty}\Delta_{1}^{(p)}(q; N)\, (u-1)^{p}\bigg)+\mc O(q^{2N}) \bigg]\, ,
\ee
with 
\be
\Delta_{1}^{(2)}(q; N) = \frac{1}{72}\, [-12(2+N)((2+N)^{2}- \E_{2}(q))-\E_{2}^{2}(q)+\E_{4}(q)]\, ,
\ee
and similar expressions for the higher functions $\Delta_{1}^{(p)}(q; N)$  that we give in explicit form and involve higher powers of Eisenstein series, as $p$ is increased.

As a further simple application of our analysis, we present novel exact predictions for the giant graviton expansion of the Schur index of a class of non-Lagrangian 
4d $\N=2$ superconformal theories with equal central charges $a=c$, introduced in \cite{Buican:2020moo,Kang:2021lic}. They are denoted $\widehat{\Gamma}(SU(N))$ with $\Gamma = D_{4}, E_{6}, E_{7}, E_{8}$,
and are built by gauging part of the flavor symmetry of a product of $D_{p}(SU(N))$ theories \cite{Cecotti:2012jx,Cecotti:2013lda}.
In these models, the unflavored Schur index is equal to certain specializations of the flavored Schur index of $\N=4$ $SU(N)$ SYM \cite{Buican:2020moo,Kang:2021lic}. \footnote{
See  \cite{Jiang:2024baj} for a recent discussion of modular properties in these models.}
This identification come from  an isomorphism between the chiral algebras of the respective theories and was explicitly
checked for the Schur index of the $\widehat{E}_{6}(SU(2))$ theory \cite{Buican:2016arp,Buican:2020moo} showing  that it is same as 
the vacuum character of the $\mc A(6)$ algebra \cite{Feigin:2007sp,Feigin:2008sg}.
To exploit this relation to derive a giant graviton expansion requires
more information than the unflavored limit in  (\ref{1.8}). From our results in flavored case  we can work out the giant graviton expansion of the $\widehat{\Gamma}(SU(N))$ theories
from the D3 brane representation  (\ref{1.4}). For instance, for the $\Gamma=E_{6}$ theory, the first two terms in its
giant graviton expansion read
\bea
\la{1.14}
& \I^{\widehat{E}_{6}(SU(N))}(q) =
\frac{(q; q^{3})_{\infty}(q^{2}; q^{3})_{\infty}}{(q)_{\infty}(q^{2})_{\infty}(q^{3})_{\infty}}
\bigg[
1+q^{N+1}\,F_{1}(q)+q^{2N+3}\, \bigg(N+F_{2}(q)\bigg) +\mc O(q^{3N})
\bigg]\, , \\
& F_{1}(q) = -\frac{(q^{2})_{\infty}^{2}}{(q; q^{2})_{\infty}^{2}} = -\frac{1}{4}\, q^{\frac{1}{4}}\,\vth_{2}(\sqrt{q})^{2}, 
\qquad F_{2}(q) = 3+\frac{1}{8}\frac{\partial^{2}_{z}\vth_{4}(0; q)}{\vth_{4}(q)}\, .
\eea
As we remarked, results like (\ref{1.14}) originate from a fruitful  combination of the D3 brane giant graviton expansion of the $\N=4$ SYM $U(N)$ Schur index
and the remarkable relation with the index of the $\widehat{\Gamma}(SU(N))$ theories. It would be  interesting to give a  gravity interpretation of these new 
brane-like expansions by exploring their large $N$ limit and associating each $q^{nN}$ term  to suitable non-perturbative corrections. 
For instance, working with truncated expansions in $q$, this was 
shown to be possible in \cite{Imamura:2021dya}
for Argyres-Douglas \cite{Argyres:1995jj,Argyres:1995xn} and  Minahan-Nemeschansky theories \cite{Minahan:1996cj,Minahan:1996fg} which are 4d $\N=2$ theories on D3-branes in 7-brane backgrounds with constant 
axio-dilaton.

\paragraph{Plan of the paper} In Section \ref{sec:fermi}, we review explicit formulas for the exact Schur index in $\N=4$ $U(N)$ SYM in terms of twisted 
Weierstrass functions. In Section \ref{sec:wrap}, we examine the corresponding expressions after application of  the analytic continuation relation (\ref{1.6}). We discuss the associated 
explicit brane indices $\I^{\rm D3}_{n}$ and their general structure at increasing $n$. In Section \ref{sec:unref}, we show how the exact unflavored  index is recovered at all orders in $q$
by presenting the quasimodular expressions of the coefficient functions in (\ref{1.11}).
Section \ref{sec:near} is devoted to the analysis of the corrections in the near-unflavored limit, \ie the expansion of the index around in powers of $u-1$.
Finally, in Section \ref{sec:gamma}, we discuss the giant graviton expansion of the $\N=2$ $\widehat{\Gamma}(SU(N))$ theories with $a=c$. Several appendices
contain technical details and additional remarks.

\section{Schur index of $\N=4$ $U(N)$ SYM from Fermi gas}
\la{sec:fermi}

To simplify some of the following expressions, let us introduce a separate notation for the Schur index of $\N=4$ $U(N)$ SYM in (\ref{1.3}) with the replacement  $q\to q^{1/2}$
\be
\la{2.1}
\widetilde \I^{U(N)}(u; q) =  \I^{U(N)}(u; q^{\frac{1}{2}})\, .
\ee
The index may be computed by the following  holonomy multiple contour integral representation (see for instance Eq.~(2.6) of \cite{Gaiotto:2019jvo})
\be
\la{2.2}
\widetilde \I^{U(N)}(u; q) = \frac{1}{N!}\frac{(q)^{2N}_{\infty}}{(q^{\frac{1}{2}}u^{\pm 1}; q)^{N}_{\infty}}\oint_{|z_{i}|=1}\prod_{i=1}^{N}\frac{dz_{i}}{2\pi i z_{i}}\,
\frac{\prod_{i\neq j}(\frac{z_{i}}{z_{j}};q)_{\infty}(q\frac{z_{i}}{z_{j}};q)_{\infty}}{\prod_{i\neq j}(q^{\frac{1}{2}}u^{-1}\frac{z_{i}}{z_{j}}; q)_{\infty}
(q^{\frac{1}{2}}u\frac{z_{i}}{z_{j}}; q)_{\infty}}\, .
\ee
The multiple integrals (\ref{2.2}) have been computed in \cite{Hatsuda:2022xdv} by Fermi gas methods. To present their result, we define
$\xi$, $\zeta$, and $\tau$ by relations
\be
\la{2.3}
\xi = u\,q^{-\frac{1}{2}} = e^{2\pi i \zeta}, \qquad q=e^{2\pi i \tau}\, , 
\ee
and introduce an auxiliary fugacity $w$ with associated chemical potential $\nu$  
\be
w=e^{2\pi i \nu}\, .
\ee
Then, the function of $\xi$ and $q$ defined as \footnote{This is simply $\widetilde\I^{U(N)}(u; q)$ with $u$ given in terms of $\xi$ in (\ref{2.3}). For later use, it will be convenient to avoid 
the common shortcut of using the same name
for a function expressed in different variables.}
\be
\la{2.5}
\bar\I^{U(N)}(\xi; q) \equiv \widetilde\I^{U(N)}(\xi\,q^{\frac{1}{2}}; q)\,,
\ee
is given by 
\be
\la{2.6}
\bar\I^{U(N)}(\xi; q)  = \frac{(-1)^{N}\, \xi^{\frac{N^{2}}{2}}\vth(w,q)}{\vth(w\xi^{-N},q)}\, \mc Z_{N}(w,\xi,q)\, .
\ee
In this expression,  the $q$-theta function $\vth$ is defined in (\ref{A.6}) and  the functions $\mc Z_{N}(w,\xi,q)$ are given by 
\be
\la{2.7}
\mc Z_{N}(w,\xi,q) = \sum_{\Lambda\vdash N}(-1)^{N-|\Lambda|}\prod_{i=1}^{r}\frac{1}{\lambda_{i}^{m_{i}}m_{i}!}Z_{\lambda_{i}}(w,\xi,q)^{m_{i}},
\ee
where as usual $\Lambda\vdash N$ means that $\Lambda$ is a partition of $N$ represented as  
\be
\la{2.8}
\Lambda = (\lambda_{1}^{m_{1}}\cdots\lambda_{r}^{m_{r}}), \qquad \sum_{i=1}^{r}m_{i}\lambda_{i}=N, \qquad  \lambda_{1}>\lambda_{2}>\cdots >\lambda_{r}>0, \qquad 
|\Lambda| = \sum_{i=1}^{r}m_{i}\, , 
\ee
\ie $\Lambda$ corresponds to a Young tableau with $m_{1}$ rows with $\lambda_{1}$ columns, \textit{etc.}. 
The quantities $Z_{\ell}$ in (\ref{2.7})  are computed explicitly by the relations 
\bea
Z_{1}(w,\xi,q) &= \wei{1}{\xi}{1}(\nu,\tau)\, , \\
Z_{\ell}(w,\xi,q) &= \frac{w^{-(\ell-1)}}{(\ell-1)!}\sum_{k=1}^{\ell-1}k!\, |s_{\ell-1,k}|\, \wei{k+1}{q^{\ell-1}\xi^{\ell}}{1}(\nu, \tau)\qquad (\ell\ge 2)\, ,
\eea
where $s_{a,b}$ are Stirling number of the first kind and $P_{k}$ are twisted Weierstrass functions defined in  (\ref{A.14}). It can be shown that the index expression (\ref{2.6})
does not depend on the choice of the auxiliary fugacity $w$ and special choices of $w$ may simplify expressions.

Let us give some explicit examples, following \cite{Hatsuda:2022xdv}, for the purpose of illustration and to  give an idea of the typical expressions. 
In the simplest $U(1)$ case, the index expressed in terms of $\xi$ and $q$ reads
\be
\bar \I^{U(1)}(\xi; q) = \widetilde\I^{U(1)}(\xi\,q^{\frac{1}{2}}; q) = -\frac{\xi^{\frac{1}{2}}\vth(w,q)}{\vth(w\xi^{-1},q)}\, \wei{1}{\xi}{1}(\nu,\tau)\, .
\ee
Using (\ref{A.14}),
we get the simple result
\be
\la{2.11}
\bar \I^{U(1)}(\xi; q) = \frac{(q)_{\infty}^{2}}{(\xi^{-1},q)_{\infty}(q\xi,q)_{\infty}} = -\sqrt\xi \frac{(q)_{\infty}^{3}}{\vth(\xi^{-1};q)}\, ,
\ee
that, in particular,  shows explicitly the independence on the auxiliary fugacity $w$.
Going back to $(u; q)$ variables, this is 
\be
\la{2.12}
\widetilde\I^{U(1)}(u; q) =\frac{(q)_{\infty}^{2}}{(q^{\frac{1}{2}}u^{\pm 1},q)_{\infty}}\, , 
\ee
whose $q$-series is 
\ba
\widetilde\I^{U(1)}(u; q) &= 1+(u+u^{-1})\, q^{\frac{1}{2}}+(u^{2}-1+u^{-1})\, q+(u^{3}+u^{-3})\, q^{\frac{3}{2}}\lp
+(u^{4}+u^{-4})\, q^{2}+(u^{5}-u-u^{-1}+u^{-5})\, q^{\frac{5}{2}}+\cdots\, ,
\ea
and of course agrees with the explicit evaluation of (\ref{2.2}).
For the $U(2)$ theory, we have 
\be
\la{2.14}
\bar\I^{U(2)}(\xi,q) = \frac{\xi^{2}\vth(w; q)}{2\vth(w\xi^{-2}; q)}\bigg[\wei{1}{\xi}{1}^{2}(\nu,\tau)-\frac{1}{w}\wei{2}{q\xi^{2}}{1}(\nu,\tau)\bigg]
\ee
To derive a $q$-expansion, it is convenient to fix the freedom in the choice of $w$ by taking $w=\xi$. This gives the simpler result \footnote{From (\ref{A.14}) we have
$\wei{1}{\xi}{1}(\zeta, \tau) =  \frac{(q)_{\infty}^{3}\vth(1,q)}{\vth(\xi^{-1},q)\vth(\xi,q)}=0$ because $\vth(1,q)\sim (1,q)_{\infty}=0$.}
\be
\la{2.15}
\bar\I^{U(2)}(\xi,q) = \frac{\xi}{2}\wei{2}{q\xi^{2}}{1}(\zeta,\tau)
\ee
We have the explicit expression of the twisted Weiserstrass function
\be
\wei{2}{x}{1}(\nu,\tau) = \frac{(q)^{3}_{\infty}\vth(x^{-1}w; q)}{\vth(x^{-1};q)\vth(w;q)}\bigg[w\frac{\vth'(w;q)}{\vth(w;q)}-x^{-1}w\frac{\vth'(x^{-1}w; q)}{\vth(x^{-1}w; q)}\bigg]\,,
\ee
and thus we can write 
\be
\la{2.17}
\bar\I^{U(2)}(\xi,q) = \frac{\xi}{2}\frac{(q)^{3}_{\infty}\vth(\frac{1}{q\xi}; q)}{\vth(\frac{1}{q\xi^{2}};q)\vth(\xi;q)}\bigg[\xi\frac{\vth'(\xi;q)}{\vth(\xi;q)}-\frac{1}{q\xi}\frac{\vth'(\frac{1}{q\xi}; q)}{\vth(\frac{1}{q\xi}; q)}\bigg]
\ee
Replacing $\xi= u q^{-\frac{1}{2}}$ and expanding in small $q$ gives same as evaluation of (\ref{2.2}) as one can check, \ie one obtains the $q$-series
\ba
\widetilde \I^{U(2)}(u; q) &= 1+(u+u^{-1})q^{\frac{1}{2}}+2(u^{2}+u^{-2})q+2(u^{3}+u^{-3})q^{\frac{3}{2}}\lp
+3(u^{4}+u^{-4})q^{2}+(3u^{5}+u+u^{-1}+3u^{-5})q^{\frac{5}{2}}+\cdots\, .
\ea

\section{Wrapped D3 brane indices from analytic continuation}
\la{sec:wrap}

The giant-graviton expansion of the $\N=4$ $U(N)$ SYM Schur index is given by the representation (\ref{1.4})
and requires the brane indices defined by the analytic continuation  (\ref{1.6}). 
We have, \cf (\ref{2.1}), 
\ba
\I_{n}^{\rm D3}(u; q) &=  \I^{U(n)}( q^{-\frac{3}{2}}u^{-\frac{1}{2}};q^{\frac{1}{2}}u^{-\frac{1}{2}}) =\widetilde \I^{U(n)}(q^{-\frac{3}{2}}u^{-\frac{1}{2}}; \tfrac{q}{u})\,,
\ea
In the right hand side the combination $\xi$ is $1/q^{2}$, \cf (\ref{2.3}),  and therefore relation (\ref{1.6}) amounts to the simple correspondence 
\be
\la{3.2}
\I_{n}^{\rm D3}(u; q) =  \bar\I^{U(n)}(\tfrac{1}{q^{2}}, \tfrac{q}{u})\,.
\ee
Let us apply systematically this transformation to the $U(n)$ indices discussed in the previous section.

\paragraph{Single wrapping}

At leading order in the giant graviton expansion, we apply (\ref{3.2}) to (\ref{2.11}) and  get immediately the expression
\ba
\la{3.3}
\I^{\rm D3}_{1}(u; q) &=  -\frac{1}{q}\frac{(\frac{q}{u})^{3}_{\infty}}{
\vth(q^{2}; \frac{q}{u})} = -qu^{2}\frac{({\frac{q}{u}})_{\infty}^{3}}{\vth(u^{2}; \frac{q}{u})}\,,
\ea
where we used (\ref{A.8}). Expanding in small $q$ gives
\ba
\la{3.4}
\I^{\rm D3}_{1}(u; q) &= \frac{u^{3}}{1-u^{2}}q+(1-u^{2})q^{2}+(u^{-3}-u^{3})q^{3}+\cdots\, .
\ea
in agreement with the result (\ref{1.7}) in \cite{Arai:2020qaj}. Eq.~(\ref{3.3}) represents its closed expression at any $q$. The above derivation is straightforward. Other 
more compact representations
of the $U(N)$ index turn out to be less convenient for analytic continuation, but will play a role in later Section  \ref{app:alt}.
  
\paragraph{Double wrapping}

At wrapping 2, \ie at subleading order in the giant graviton expansion, we apply (\ref{3.2}) to (\ref{2.17}) and get the expression
\be
\I^{\rm D3}_{2}(u;q) =  
 \frac{1}{2q^{2}}\frac{(\frac{q}{u})^{3}_{\infty}\vth(qu; \frac{q}{u})}{\vth(q^{3}u;\frac{q}{u})\vth(\frac{1}{q^{2}};\frac{q}{u})}\bigg[\frac{1}{q^{2}}\frac{\vth'(\frac{1}{q^{2}};\frac{q}{u})}
 {\vth(\frac{1}{q^{2}};\frac{q}{u})}-qu\frac{\vth'(qu; \frac{q}{u})}{\vth(qu; \frac{q}{u})}\bigg]\,.
\ee
We can use (from (\ref{A.7})) the relations
\be
\vth(x;q) = -\vth(x^{-1};q), \qquad \frac{\vth'(x;q)}{\vth(x;q)} = -\frac{1}{x^{2}}\frac{\vth'(x^{-1};q)}{\vth(x^{-1};q)}\, , 
\ee
to simplify it into 
\be
\la{3.7}
\I^{\rm D3}_{2}(u;q) =  
 \frac{1}{2q}\frac{(\frac{q}{u})^{3}_{\infty}\vth(qu; \frac{q}{u})}{
 \vth(q^{3}u;\frac{q}{u})\vth(q^{2};\frac{q}{u})}\bigg[q\frac{\vth'(q^{2};\frac{q}{u})}
 {\vth(q^{2};\frac{q}{u})}+u\frac{\vth'(qu; \frac{q}{u})}{\vth(qu; \frac{q}{u})}\bigg]\,.
\ee
We cannot naively expand the ratios $\vth'/\vth=\partial_{x}\log \vth$ inside bracket by first expanding $\vth$ when the first argument of the $\vth$ function 
has an explicit integer power of $q$. However, using systematically (\ref{A.8}) and (\ref{A.10}), we get bring (\ref{3.7}) to the form  
\ba
\I^{\rm D3}_{2}(u;q) &=  
 \frac{q^{4}u^{8}}{2}\frac{(\frac{q}{u})^{3}_{\infty}}{\vth(u^{4};\frac{q}{u})}
 \bigg[2u^{2}\frac{\vth'(u^{2},\frac{q}{u})}{\vth(u^{2},\frac{q}{u})}-3\bigg]\,.
 \ea
Expanding this expression at small $q$ gives the series 
\ba
\la{3.9}
\I^{\rm D3}_{2}(u;q) & = \frac{u^{10} (2-u^2) q^4}{(1-u^2)(1-u^{4})}+u^5 (2-u^4) q^5+(2+2 
u^6-u^{12}) q^6+(2u^{-5}+3 u^7-u^{15}) 
q^7\lp
+(3+2u^{-10}+3 u^8-u^{18}) 
q^8+(2u^{-15}+u^{-3}+4 u^9-u^{21}) q^9+\cdots \, ,
\ea
which is again in agreement with (\ref{1.7}) and extends it to all orders in $q$.

\paragraph{Higher wrapping}

The same procedure can be repeated at higher wrapping without difficulty, because the steps we illustrated for the $U(1)$ and $U(2)$ cases 
are fully algorithmic and easily coded.
 To present the results in compact form 
it is convenient to define the quantities 
\be
Q=(\tfrac{q}{u})_{\infty}, \qquad R_{k,p}= R_{k}(u^{p},\tfrac{q}{u}), \qquad \Theta_{p}= \vth(u^{p},\tfrac{q}{u})\, ,
\ee
where the function $R_{k}(x; q)$ is defined in (\ref{A.9}). At wrapping 3,4,5, we obtain 
the brane indices
\ba
\I^{\rm D3}_{3}(u;q) &= -\frac{q^9 u^{14}Q^{3}}{6\Theta_{6}}[ 20 u^4+2 
R_{1,-2}^2-8 u^8 R_{1,4}+R_{1,-2} (8 u^2-2 u^6 
R_{1,4})-R_{2,-2}+u^{12} R_{2,4}]\, , \\
\I^{\rm D3}_{4}(u;q) &= \frac{1}{24} q^{16} Q^3 u^{26} \bigg[\frac{3 Q^3 u^2 \Theta _2^2 (3 
u^2+R_{1,-2}-u^4 R_{1,2}){}^2}{\Theta _{-2} \Theta _4^2 \Theta_6}+\frac{1}{\Theta_{8}}[-210 u^6-6 R_{1,-2}^3\lp
+6 u^2 R_{1,-2}^2 (-5+u^6 R_{1,6})+15 
u^2 R_{2,-2}+R_{1,6} (90 u^{12}-3 u^8 R_{2,-2})-15 u^{18} 
R_{2,6}\lp
+R_{1,-2} (-90 u^4+30 u^{10} R_{1,6}+6 R_{2,-2}-3 u^{16} 
R_{2,6})-R_{3,-2}+u^{24} R_{3,6}]\bigg]\, , \\
\I^{\rm D3}_{5}(u;q) &= \frac{1}{120} q^{25} Q^3 u^{42} \bigg[-\frac{10 Q^3 u^2 \Theta _2}{\Theta _{-2} \Theta_6 \Theta _8}[(-3 
u^2-R_{1,-2}+u^4 R_{1,2}) (20 u^4+2 R_{1,-2}^2-8 u^8 R_{1,4}\lp
+R_{1,-2} 
(8 u^2-2 u^6 R_{1,4})-R_{2,-2}+u^{12} R_{2,4})]+\frac{1}{\Theta_{10}}[-3024 u^8-24 R_{1,-2}^4\lp
+24 u^2 R_{1,-2}^3 (-6+u^8 \
R_{1,8})+252 u^4 R_{2,-2}-6 R_{2,-2}^2-252 u^{24} R_{2,8}+6 u^{20} 
R_{2,-2} R_{2,8}\lp
-12 R_{1,-2}^2 (42 u^4-12 u^{12} R_{1,8}-3 
R_{2,-2}+u^{20} R_{2,8})-24 u^2 R_{3,-2}\lp
+4 u^{10} R_{1,8} (336 u^6-18 
u^2 R_{2,-2}+R_{3,-2})+24 u^{32} R_{3,8}\lp
+4 R_{1,-2} (-336 u^6+6 
u^{10} R_{1,8} (21 u^4-R_{2,-2})+36 u^2 R_{2,-2}-18 u^{22} R_{2,8}-2 
R_{3,-2}+u^{30} R_{3,8})\lp
+R_{4,-2}-u^{40} R_{4,8}]\bigg]\, .
\ea
Although the expressions have increasing complexity, their expansion in powers of $q$ is straightforward and one finds 
(we write only the first terms for brevity, but it is clear that the expansions can be lengthen at will)
\ba
\la{314}
\I_{3}^{\rm D3} & (u;q) = \frac{u^{21} (5-3 u^2-3 u^4+2 u^6)}{(1-u^2) (1-u^4) 
(1-u^6)}\,q^{9}+\frac{u^{14} (5-3 u^4-3 u^6+2 
u^{10})}{1-u^4}\,q^{10}\lp
+\frac{u^7 (5-5 u^4+2 u^6+4 u^8-2 
u^{14}-u^{16}-2 u^{18}+2 u^{22})}{1-u^4}\,q^{11} +\cdots\, , \\
\I_{4}^{\rm D3}& (u;q) = -\frac{ u^{36} (-14+9 u^2+10 u^4+2 u^6-6 u^8-7 u^{10}+5 
u^{12})}{(1-u^2) (1-u^4) (1-u^6) (1-u^8)}\,q^{16}\lp
-\frac{u^{27} (-14+9 
u^4+9 u^6+10 u^8-6 u^{10}-7 u^{12}-7 u^{14}+5 u^{18})}{(1-u^4) 
(1-u^6)}\,q^{17}\lp
-\frac{u^{18} (-14+14 u^2+14 u^4-19 u^6+2 
u^{12}+u^{14}+u^{16}+2 u^{18}+3 u^{24}-5 u^{26}-5 u^{28}+5 
u^{30})}{(1-u^2) (1-u^4)}\,q^{18} +\cdots\, , \\
\I_{5}^{\rm D3}& (u;q) = \frac{ u^{55} (42-28 u^2-32 u^4-9 u^6-6 u^8+45 u^{10}+3 
u^{12}+7 u^{14}-16 u^{16}-19 u^{18}+14 u^{20})}{(1-u^2) (1-u^4) 
(1-u^6) (1-u^8) (1-u^{10})}\,q^{25}\lp
+\frac{u^{44}}{(1-u^4) (1-u^6) (1-u^8)} (42-28 u^4-28 u^6-32 
u^8-9 u^{10}+15 u^{12}+45 u^{14}+29 u^{16}+7 u^{18}\lp
-16 u^{20}-19 
u^{22}-19 u^{24}+14 u^{28})\,q^{26}+\frac{ 
u^{33}}{(1-u^4) (1-u^6) (1-u^8)} (42-42 u^4-28 u^6-28 u^8\lp
+42 u^{10}+52 u^{12}+15 u^{14}-13 
u^{16}-38 u^{18}-8 u^{20}-9 u^{22}-3 u^{24}+2 u^{26}+3 u^{28}-11 
u^{30}\lp
+10 u^{32}+29 u^{34}+12 u^{36}+14 u^{38}-19 u^{40}-19 u^{42}-14 
u^{44}+14 u^{48})\,q^{27}+\cdots\, .
\ea
Continuing this way, the general structure of $\I^{\rm D3}_{n}(u;q)$ turns out to be 
\ba
\la{3.17}
\I^{\rm D3}_{n}(u; q) &= \frac{u^{n(2n+1)}A_{n}(u)}{\prod_{k=1}^{n}(1-u^{2k})}q^{n^{2}}+\frac{B_{n}(u; q)}{(1-u^{2})^{n-2}}\,q^{n^{2}+1} \, ,
\ea
where $A_{n}(u)$ are  polynomial in $u$ with degree $n(n-1)$ and $B_{n}(u; q)$ has a $q$-series with coefficients that are rational functions of $u$, smooth for $u\to 1$. The first cases of the polynomials $A_{n}(u)$ are 
\ba
\la{3.18}
A_{1}(u) &= 1, \qquad A_{2}(u) = 2-u^{2}, \qquad A_{3}(u) = 5-3 u^2-3 u^4+2 u^6, \notag \\
A_{4}(u) &= 14-9 u^2-10 u^4-2 u^6+6 u^8+7 u^{10}-5 u^{12}, \notag \\
A_{5}(u) &= 42-28 u^2-32 u^4-9 u^6-6 u^8+45 u^{10}+3 u^{12}+7 u^{14}-16 u^{16}-19 
u^{18}+14 u^{20}, \notag \\
A_{6}(u) &= 132-90 u^2-104 u^4-31 u^6-29 u^8+71 u^{10}+80 u^{12}+99 u^{14}-41 
u^{16}-52 u^{18}\\
& -60 u^{20}-15 u^{22}-20 u^{24}+47 u^{26}+56 u^{28}-42 
u^{30}, \notag \\
A_{7}(u) & =429-297 u^2-345 u^4-104 u^6-104 u^8+210 u^{10}+12 u^{12}+582 
u^{14}+102 u^{16}-103 u^{18}\notag \\
& -188 u^{20}-201 u^{22}-309 u^{24}-83 
u^{26}+328 u^{28}-50 u^{30}+203 u^{32}+47 u^{34}+60 u^{36}-146 
u^{38}\notag\\
& -174 u^{40}+132 u^{42}, \notag\\
A_{8}(u) &= 1430-1001 u^2-1166 u^4-351 u^6-359 u^8+695 u^{10}-64 u^{12}+1149 
u^{14}+1177 u^{16}\notag\\
& +497 u^{18}-427 u^{20}-524 u^{22}-1715 u^{24}-344 
u^{26}-318 u^{28}-507 u^{30}+1076 u^{32}+682 u^{34}\notag\\
& +826 u^{36}+304 
u^{38}-398 u^{40}-361 u^{42}+86 u^{44}-652 u^{46}-149 u^{48}-188 
u^{50}\notag\\
& +471 u^{52}+561 u^{54}-429 u^{56}\, ,\notag
\ea
and one can check that in all cases $A_{n}(1)=1$ and coefficients are integer. 
An explicit formula for them  is illustrated in next section, see in particular formula (\ref{B22}).
The functions $B_{n}(u; q)$ are non-trivial, even at a fixed value of $u$, having a full expansion in powers of $q$, and will be discussed later on.

\subsection{Other representations of the $U(N)$ index and polynomials $A_{N}(u)$}
\la{app:alt}

We now show  how to get an explicit formula for the polynomials $A_{N}(u)$ in (\ref{3.17}).
We start from an alternative compact representation of the indices $\bar\I^{U(N)}(\xi; q)$ derived in  \cite{Hatsuda:2022xdv} 
\be
\la{B1}
\bar \I^{U(N)}(\xi; q) = -\mathop{\sum_{p_{1}, \cdots, p_{N}\in \mathbb{Z}}}_{p_{1}<\cdots < p_{N}}\xi^{\frac{N^{2}}{2}}\prod_{n=1}^{N}\frac{\xi^{-p_{n}}}{1-\xi^{\frac{N}{2}}q^{p_{n}}}\,.
\ee
In the simplest $U(1)$ case, it is 
\be
\bar\I^{U(1)}(\xi;q) = -\sum_{p\in\mathbb Z}\frac{\xi^{-p+\frac{1}{2}}}{1-\xi^{\frac{1}{2}}q^{p}}\,,
\ee
and the associated brane index is obtained from relation (\ref{3.2}). Applying it to the summand, we get 
\be
\I_{1}^{\rm D3}(u;q)  = -\sum_{p\in\mathbb Z}\frac{q^{2p-1}}{1-u^{-p}q^{p-1}}\,.
\ee
We now split the sum with positive and negative indices and write
\ba
\I_{1}^{\rm D3}(u;q) &= -\sum_{p=0}^{\infty}\frac{q^{2p-1}}{1-u^{-p}q^{p-1}}-\sum_{p=1}^{\infty}\frac{q^{-2p-1}}{1-u^{p}q^{-p-1}} = 
-\sum_{p=0}^{\infty}\frac{q^{2p-1}}{1-u^{-p}q^{p-1}}+\sum_{p=1}^{\infty}\frac{u^{-p}q^{-p}}{1-u^{-p}q^{p+1}}\,.
\ea
In the last term, we expand the denominator and exchange the order of the two sums
\be
\sum_{p=1}^{\infty}\frac{u^{-p}q^{-p}}{1-u^{-p}q^{p+1}} = \sum_{p=1}^{\infty}\sum_{n=0}^{\infty}u^{-p}q^{-p}u^{-pn}q^{n(p+1)} = -\sum_{n=0}^{\infty}\frac{q^{2n-1}}{q^{n-1}-u^{n+1}}\,.
\ee
Thus, we can write
\be
\I_{1}^{\rm D3}(u;q) = \sum_{p=0}^{\infty}\frac{1}{u^{2}}\left(\frac{q}{u}\right)^{2p-1}\frac{u^{p}(1-u^{p+1})-q^{p-1}(1-u^{p})}{(1-u^{-p}q^{p-1})(1-u^{-p-1}q^{p-1})}
\ee
This expression can be expanded at small $q$ with the result
\ba
F(u;q) &= \frac{u^3 q}{1-u^2}+(1-u^2) q^2+(u^{-3}-u^3) 
q^3+(1+u^{-6}-u^{-2}-u^4) q^4+(u^{-9}-u^5) 
q^5+\cdots\, ,
\ea
in agreement with (\ref{3.4}). 
Extending this procedure to higher $N$  by the same methods provides
a way to compute  the leading term $\sim q^{n^{2}}$ in $\I^{\rm D3}_{n}(u; q)$, \ie the polynomials $A_{n}(u)$ in (\ref{3.18}).
To show this, we  begin with the algebraic identity, following from Polya theory \cite{stanley2023enumerative},  (\cf (\ref{2.8}) for notation of 
partitions)
\be
\la{B8}
\sum_{0\le i_{1}<\cdots < i_{k}}x^{i_{1}+\cdots +i_{k}} = 
\frac{x^{k(k-1)/2}}{(1-x)(1-x^{2})\cdots(1-x^{k})} = \sum_{\Lambda \vdash k}
(-1)^{k+|\Lambda|}\prod_{i=1}^{p}\frac{1}{\lambda_{i}^{m_{i}}(m_{i}!)}\frac{1}{(1-x^{\lambda_{i}})^{m_{i}}}\,.
\ee
Eq.~(\ref{B8}) determines the coefficients in the decomposition of the total trace of a symmetric tensor with summation restricted to ordered indices 
in terms of unrestricted traces. For a symmetric tensor $T_{i_{1}, \dots, i_{k}}$ we have from (\ref{B8})
\be
\la{B10}
\sum_{i_{1}<\cdots < i_{k}}T_{i_{1}, \dots, i_{k}} = \sum_{\Lambda \vdash k}
\frac{(-1)^{k+|\Lambda|}}{\prod_{q=1}^{p}\lambda_{q}^{m_{q}}(m_{q}!)}
\sum_{j_{1}, \dots, j_{|\Lambda|}}\ T_{J}\, ,
\ee
where $J$ is a multi-index with, for all $p$,  $m_{p}$ groups of the same index repeated $\lambda_{p}$ times. The number of free indices is  $|\Lambda|$.
For instance, for a symmetric tensor with rank 3 or 4, the expansion (\ref{B10}) gives the general identities
\ba
\sum_{i<j<k}T_{ijk} &= \frac{1}{6}\sum_{i,j,k}T_{ijk}-\frac{1}{2}\sum_{i,j}T_{ijj}+\frac{1}{3}\sum_{i}T_{iii}, \\
\sum_{i<j<k<\ell}T_{ijk\ell} &= \frac{1}{24}\sum_{i,j,k,\ell}T_{ijk\ell}-\frac{1}{4}\sum_{i,j,k}T_{ijkk}+\frac{1}{8}\sum_{i,j}T_{iijj}
+\frac{1}{3}\sum_{i,j}T_{ijjj}-\frac{1}{4}\sum_{i}T_{iiii}\, .
\ea
Considering now the symmetric rank $N$ tensor in (\ref{B1}) 
\be
T_{p_{1}, \dots, p_{N}} = -\prod_{n=1}^{N}\frac{\xi^{-p_{n}+\frac{N}{2}}}{1-\xi^{\frac{N}{2}}q^{p_{n}}}\,,
\ee
we obtain from (\ref{B10})
\be
\la{B14}
\I^{\rm D3}_{N}(u; q) = \sum_{\lambda \vdash N}
(-1)^{N+|\lambda|+1}\prod_{q=1}^{p}\frac{1}{\lambda_{q}^{m_{q}}(m_{q}!)} S^{(N)}_{\lambda_{q}}(u; q)^{m_{q}}.
\ee
The functions $S^{(N)}_{r}(u; q)$ are obtained by applying the  procedure we adopted to treat the $U(1)$ case to the single index sums. 
Their explicit expression is 
\ba
\la{B15}
S^{(N)}_{r}(u; q) = \sum_{n=0}^{\infty}\bigg[\bigg(\frac{q^{2n-N}}{1-u^{-n}q^{n-N}}\bigg)^{r}
+(-1)^{r}\binom{n}{r-1}\,\frac{u^{-n-1}q^{(n+1)(N+1)-r(N+2)}}{1-u^{-n-1}q^{n-2r+1}}
\bigg]\, ,
\ea
where we used
\be
\frac{1}{(1-x)^{r}} =\sum_{n=0}^{\infty} \binom{n}{r-1}\,x^{n-r+1}\, .
\ee
To give an example, for $N=3$, we have 
\be
\la{B17}
I^{\rm D3}_{3}(u; q) = -\frac{1}{6}(S^{(3)}_{1})^{3}+\frac{1}{2}S^{(3)}_{1}S^{(3)}_{2}-\frac{1}{3}S^{(3)}_{3}\, ,
\ee
with 
\ba
S^{(3)}_{1}(u;q) &= \frac{u^8 q^3}{1+u-u^3-u^4}+(u^4-u^6) q^4+(1-u^3+u^5-u^8) 
q^5+\bigg(-1+\frac{1}{u^4}+u^6-u^{10}\bigg) 
q^6 \lp
+\bigg(1+\frac{1}{u^8}-\frac{1}{u^3}-u^4+u^7-u^{12}\bigg) 
q^7+\bigg(\frac{1}{u^{12}}-\frac{1}{u^6}+u^8-u^{14}\bigg) 
q^8\lp
+\bigg(\frac{1}{u^{16}}-\frac{1}{u^9}+\frac{1}{u^5}-u^5+u^9-u^{16}\bigg) q^9+\cdots, \\
S^{(3)}_{2}(u;q) &= \frac{u^{15} (4+u+u^2-2 u^3) q^6}{(-1+u^3)^2 (1+u+u^2+u^3)}-2 (u^{10} 
(-2+u^2)) q^7-u^5 (-4+u^3-5 u^7+2 u^{10}) q^8\lp
+(4+6 u^{14}-2 u^{18}) 
q^9+\cdots, \\
S^{(3)}_{3}(u;q) &= \frac{u^{21} (15-19 u^3+6 u^6) q^9}{(-1+u^3)^3 (1+u^3)}+\cdots\, ,
\ea
and from (\ref{B17}), we get 
\be
\I^{\rm D3}_{3}(u; q) = \frac{u^{21} (5-3 u^2-3 u^4+2 u^6)}{(1-u^2) (1-u^4) 
(1-u^6)}\,q^{9}+\cdots, 
\ee
in agreement with (\ref{314}). The fact that $\I^{\rm D3}_{N}(u; q)$ starts at order $q^{N^{2}}$ is consistent with  
 $S^{(N)}_{r}(u; q) = C^{(N)}_{r}(u)\,q^{Nr}+\mc O(q^{Nr+1})$ and therefore $A_{N}(u)$ in (\ref{3.18}) can be read from (\ref{B14}) that gives
\ba
\la{B22}
& A_{N}(u) = u^{-N(2N+1)}\ \prod_{k=1}^{N}(1-u^{2k})\ \sum_{\lambda \vdash N}
(-1)^{N+|\lambda|+1}\prod_{q=1}^{p}\frac{1}{\lambda_{q}^{m_{q}}(m_{q}!)} [C^{(N)}_{\lambda_{q}}(u)]^{m_{q}}\,, \\
& C^{(N)}_{r}(u) = (-1)^{r}\, u^{rN}\,\bigg[(1-u^{N})^{-r}-\binom{2r-1}{r-1}\frac{u^{rN}}{1-u^{2r}}\bigg]
-(-1)^{r}u^{N}\sum_{n=0}^{2r-2}\binom{n}{r-1}u^{nN}\, .\notag 
\ea
where the \underline{finite sum} representation of  $C^{(N)}_{r}(u)$ may be find after some manipulation of  (\ref{B15}). Again, let us give an example
by considering the case $N=4$. From the formula in second line of (\ref{B22}), we get 
\begin{alignat}{2}
&C^{(4)}_{1}(u) = \frac{u^{10}}{1-u^{4}} , && C^{(4)}_{2}(u) = \frac{u^{20}}{(1-u^{4})^{2}}, \\
& C^{(4)}_{3}(u) =-\frac{u^{28} (15+5 u^2-19 u^4-4 u^6+6 u^8)}{(1+u^2) (1-u^4)^2 
(1-u^6)},\ \  &&
C^{(4)}_{4}(u) = -\frac{u^{36} (-56+119 u^4-85 u^8+20 u^{12})}{(1-u^4)^3 (1-u^8)}\,, \notag
\end{alignat}
and the first line of (\ref{B22}) gives
\be
A_{4}(u) = 14 - 9 u^2 - 10 u^4 - 2 u^6 + 6 u^8 + 7 u^{10} - 5 u^{12}\,,
\ee
in agreement with (\ref{3.18}). The representation (\ref{B22}) is quite efficient and for instance we can get $A_{N}(u)$ for $N=20$ in a few seconds.

\section{Giant-graviton expansion of the index in unflavored limit}
\la{sec:unref}

The giant graviton expansion of the  Schur index  in (\ref{1.4}) and can be regrouped as  \footnote{Notice that in the l.h.s. we have $\I$ and not $\widetilde \I$.}
\ba
\la{4.1}
 \frac{\I^{U(N)}(u; q)}{\I^{\rm KK}(u;q)} &=  \sum_{n=0}^{\infty}q^{nN}\, \Isp_{n}^{\rm D3}(u; q)\, , \\
\la{4.2}
\Isp^{\rm D3}_{n}(u; q) &= \sum_{p=0}^{n} q^{2(n-p)p}\,u^{(n-2p)N}\,\I^{\rm D3}_{n-p}(u; q)\, \I^{\rm D3}_{p}(u^{-1}; q)\,.
\ea
The exact result in the unflavored limit $u\to 1$  was derived in \cite{Bourdier:2015wda} and already given in (\ref{1.8}).
Comparing it with (\ref{4.1}) gives the prediction 
\be
\la{4.3}
\lim_{u\to 1}\Isp_{n}^{\rm D3}(u; q) = (-1)^{n}\frac{(2n+N)(N+n-1)!}{n!\, N!}\,q^{n^{2}}\, ,
\ee
\ie  the following polynomials in $N$ times $q^{n^{2}}$
\bea
\la{4.4}
 \Isp_{1}^{\rm D3}(1; q) &= -(N+2)\,q, \qquad \Isp_{2}^{\rm D3}(1; q) = \frac{1}{2}(1+N)(4+N)\,q^{4}, \\
\Isp_{3}^{\rm D3}(1; q) &= -\frac{1}{6}(1+N)(2+N)(6+N)\,q^{9}, \\ 
\Isp_{4}^{\rm D3}(1; q) &= \frac{1}{24}(1+N)(2+N)(3+N)(8+N)\,q^{16}, \cdots\, .
\eea
As discussed in \cite{Beccaria:2023cuo,Beccaria:2024vfx}, the $N$ dependence of the giant graviton corrections is a familiar fact working in unrefined limits and comes from 
the zero modes of fluctuations of fields in the theory on the brane world-volume. In particular, the $N$ term in the leading single wrapping contribution is associated with 
a total of two zero modes from both scalar fields and fermion fields fluctuations in a $\N=4$ Maxwell multiplet \cite{Beccaria:2024vfx}.

\subsection{Differential constraints}

It is non-trivial to match the brane index prediction (\ref{4.3}) starting from its structure (\ref{3.17}). 
To see this in full details, let us begin with the case $n=1$, \ie at 
leading wrapping order.  From (\ref{4.2}) we get 
\be
\Isp_{1}^{\rm D3}(u; q) = -[N(A_{1}(1)+A_{1}'(1))+2A_{1}(1)]\, q+\mc O(u-1)\, .
\ee
This is in agreement with (\ref{4.4}) if 
\be
\la{4.6}
A_{1}(1)=1, \qquad A_{1}'(1) = 0\, .
\ee
These conditions are trivially true since $A_{1}(u)=1$, \cf (\ref{3.18}).
The next case is $n=2$. Using (\ref{4.6}), we get now
\ba
\la{4.7}
\Isp_{2}^{\rm D3}(u; q) &= -\frac{q^{4}}{4}(1-A_{2}(1))\bigg[\frac{1}{(u-1)^{2}}+\frac{1}{u-1}\bigg]\lp
-\frac{q^{4}}{16}\bigg[
-1-95 A_2(1)+32 q B_1(1,q)-32 q B_2(1,q)-30 A_2'(1)+4 A_1''(1)-2 
A_2''(1)\lp
-8 \big(7 A_2(1)+A_2'(1)\big) N-8 A_2(1) N^2
\bigg]+\mc O(u-1)\, .
\ea
Cancellation of the poles at $u=1$ requires $A_{2}(1)=1$ that holds, as we mentioned.  We can use the known values of  polynomials $A_{k}(u)$ and their derivatives at $u=1$. 
In fact,  as we have seen, the list of polynomials in (\ref{3.18}) can be 
computed by (\ref{B22}) in a finite number of steps when focusing on a definite wrapping order. This simplifies (\ref{4.7}) to  
\ba
\Isp_{2}^{\rm D3}(u; q) &= \frac{1}{2}q^{4}\bigg[(1+N)(4+N)-4q\,\big(B_{1}(1,q)-B_{2}(1,q)\big)\bigg]+\mc O(u-1)\, .
\ea
Comparing with (\ref{4.4}) and denoting $B_{k}(q) \equiv B_{k}(1; q)$,  we need the non-trivial condition 
\be
\la{4.9}
B_{1}(q) = B_{2}(q)\, .
\ee
Similarly, considering higher wrapping contributions, we find additional differential conditions involving the $B_{k}(q)$ functions. 
Again, these constraints come from requiring cancellation of poles at $u=1$ and 
matching powers of $N$ in the finite part. At wrapping 3, we get two conditions (we omit the $q$ argument)
\bea
\la{4.10}
& B_1+2 B_2-6 B_3 = 0, \\
& 4 B_1-4 B_2+2 B_3-B_1^{(1)}+2 B_2^{(1)}-2 B_3^{(1)} = 0\, ,
\eea
where a notation has been introduced for the partial derivatives with respect to $u$ evaluated at the point $u=1$ \footnote{We remark that $B^{(1)}(q) \neq B'(q)$.}
\be
\la{411}
B^{(p)}_{k}(q) = \partial_{u}^{p}B_{k}(u; q)\big|_{u=1}\, ,
\ee
not to be confused with derivatives of $B_{k}(q)$ with respect to $q$.
At wrapping 4, we get the four conditions
\bea
\la{4.12}
& -B_1+3 B_2-6 B_3+6 B_4 = 0, \\
& -B_1-6 B_3+24 B_4=0, \\
& -8 B_1+18 B_3-24 B_4+B_1^{(1)}-6 B_3^{(1)}+12 
B_4^{(1)} = 0, \\
& -215 B_1+279 B_2+144 q B_2{}^2-306 B_3-288 q B_1 
B_3+126 B_4+90 B_1^{(1)}-162 B_2^{(1)}\\
& \qquad +180 B_3^{(1)}-108 B_4^{(1)}-6 B_1^{(2)}+18 B_2^{(2)}-36 
B_3^{(2)}+36 B_4^{(2)} = 0\, .
\eea
Finally, at wrapping 5, there are six new  constraints  
\bea
\la{4.13}
& 3 B_1-4 B_2-12 B_3+72 B_4-120 B_5 = 0, \\
& 27 B_1-4 B_2-12 B_3+648 B_4-3000 B_5 = 0, \\
& 13 B_1-36 B_2+72 B_3-96 B_4+72 B_5-B_1^{(1)}+4 B_2^{(1)}-12 B_3^{(1)}+24 B_4^{(1)}-24 B_5^{(1)} = 0, \\
& 117 B_1-36 B_2+72 B_3-864 B_4 +1800 B_5-9 B_1^{(1)}+4 B_2^{(1)}-12 
B_3^{(1)}+216 B_4^{(1)}-600 B_5^{(1)} = 0, \\
& 669 B_1-420 B_2-628 B_3-384 
q B_2 B_3+2184 B_4+1152 q B_1 B_4-1960 B_5-150 B_1^{(1)}+136 B_2^{(1)}\\
&\qquad  +264 B_3^{(1)} 
 -1008 B_4^{(1)}+1200 B_5^{(1)}+6 B_1^{(2)}-8 B_2^{(2)}-24 B_3^{(2)}+144 B_4^{(2)}-240 
B_5^{(2)} = 0,\\
& 485 B_1+324 B_2-216 B_3+1152 q B_2 B_3-2592 B_4-3456 q B_1 B_4+936 
B_5-593 B_1^{(1)} \\
& \qquad -1152 q B_4 B_1^{(1)} 
 +1052 B_2^{(1)}+1152 q B_3 B_2^{(1)}-1476 B_3^{(1)}-1152 
q B_2 B_3^{(1)}+1656 B_4^{(1)} \\ 
& \qquad +1152 q B_1 B_4^{(1)}-792 B_5^{(1)}
 +72 B_1^{(2)}-192 B_2^{(2)}+360 
B_3^{(2)}-432 B_4^{(2)}+288 B_5^{(2)} \\ 
& \qquad -2 B_1{}^{(3)}+8 B_2{}^{(3)}-24 
B_3{}^{(3)} +48 B_4{}^{(3)}-48 B_5{}^{(3)}  = 0. 
\eea
The same pattern  continues at higher wrapping larger than 5. We will stop at this order for the sake of presentation.

\subsection{Quasimodular expansions and solution of the constraints}

To prove that the differential constraints (\ref{4.9}, \ref{4.10}, \ref{4.12}, \ref{4.13}) are satisfied at all orders in $q$ we need explicit closed expressions for the 
functions $B_{n}^{(p)}(q)$. Let us begin with $n=1$. The brane index was given in (\ref{3.3}) and we can expose its singularity at $u\to 1$ by writing
\ba
\la{4.14}
\I^{\rm D3}_{1}(u;q) &= qu^{3}\frac{(\frac{q}{u})_{\infty}^{2}}{(\frac{q}{u^{3}};\frac{q}{u})_{\infty}(u^{2}; \frac{q}{u})_{\infty}} = qu^{3}\prod_{k=0}^{\infty}\frac{[1-(\frac{q}{u})^{k+1}]^{2}}{(1-\frac{q^{k+1}}{u^{k+3}})(1-\frac{q^{k}}{u^{k-2}})}\lp
 = \frac{u^{3}}{1-u^{2}}q \prod_{k=0}^{\infty}\frac{[1-(\frac{q}{u})^{k+1}]^{2}}{(1-\frac{q^{k+1}}{u^{k+3}})(1-\frac{q^{k+1}}{u^{k-1}})}\, .
 \ea
 We now expand the logarithm of the second factor around $u=1$ obtaining
 \ba
 \la{4.15}
 \log\prod_{k=1}^{\infty} & \frac{(1-\frac{q^{k}}{u^{k}})^{2}}{(1-\frac{q^{k}}{u^{k+2}})(1-\frac{q^{k}}{u^{k-2}})} = S_{2}(q)\,(u-1)^{2}
 +S_{3}(q)\,(u-1)^{3}+S_{4}(q)\,(u-1)^{4}+\cdots\, , 
 \ea
 with the following infinite sums
 \ba
 S_{2}(q) &= \sum_{k=1}^{\infty}  \frac{4q^{k}}{(1-q^{k})^{2}} , \\
\la{4.17}
 S_{3}(q) &= -\sum_{k=1}^{\infty}  \frac{4(1+k+(k-1) q^{k})q^{k}}{(1-q^{k})^{3}}, \\ 
S_{4}(q) &=\sum_{k=1}^{\infty} \frac{[5+6k+2k^{2}-2(1-4k^{2})q^{k}+(5-6k+2k^{2})q^{2k}]q^{k}}{(1-q^{k})^{4}}\, .
 \ea
These sums can be expressed in  terms of the Eisenstein series, \cf (\ref{A.11}), as illustrated in details in Appendix \ref{app:eis}. One obtains ($\E_{2n}\equiv \E_{2n}(q)$)
\ba
S_{2}(q) &= \frac{1}{6}(1-\E_{2}), \\
\la{4.20}
S_{3}(q) &= -\frac{1}{6}(1-\E_{2})+\frac{1}{72}(\E_{2}^{2}-\E_{4}), \\
\la{4.21}
S_{4}(q) &= \frac{53}{360}-\frac{11}{72}\E_{2}-\frac{1}{48}\E_{2}^{2} 
+\frac{19}{720}\E_{4}-\frac{1}{864}\E_{2}^{3}+\frac{1}{288}\E_{2}\E_{4}-\frac{1}{432}\E_{6}\, ,
\ea
and so on. At each order we need to include a generic combination of Eisenstein series with increased total degree. In other words, the sums $S_{n}(q)$ are quasimodular forms, \ie 
polynomials in the Eisenstein series $\E_{2}, \E_{4}, \E_{6}$.
Exponentiating (\ref{4.15}) and replacing into (\ref{4.14}) gives the functions $B_{1}^{(p)}(q)$ that have the explicit quasimodular expressions
\bea
\la{4.22}
q\, B_{1}(q) &= \frac{1}{24}(1-\E_{2}),\\
q\,B_{1}^{(1)}(q) &= \frac{1}{24}(1-\E_{2})+\frac{1}{288}(\E_{2}^{2}-\E_{4}), \\
q\,B_{1}^{(2)}(q)  &= -\frac{17}{720}+\frac{1}{72}\E_{2}+\frac{1}{96}\E_{2}^{2}-\frac{1}{1440}\E_{4}-\frac{1}{1728}\E_{2}^{3}+\frac{1}{576}\E_{2}\E_{4}-\frac{1}{864}\E_{6}, \\
q\,B_{1}^{(3)}(q) &= -\frac{175}{512}+\frac{3229}{8064}\E_{2}-\frac{251}{2304}\E_{2}^{2}+\frac{55}{1152}\E_{4}+\frac{23}{3456}\E_{2}^{3}
-\frac{11}{960}\E_{2}\E_{4}+\frac{1}{135}\E_{6}\\
& -\frac{1}{4608}\E_{2}^{4}+\frac{1}{2016}\E_{2}\E_{6}-\frac{5}{16128}\E_{4}^{2}\, .
\eea
We remark that the expansion of  $B_{1}^{(p)}$ for any $p$ can be obtained systematically 
from a differential equation  discussed in Appendix \ref{C.2}. This proves that the same structure is found at each $p$. 
Still, for practical evaluation, it is  more efficient to assume a quasimodular Ansatz  and fix 
coefficients from the first terms of its $q$-series. 

For the functions $B_{2}^{(p)}(q)$, $B_{3}^{(p)}(q)$, $B_{4}^{(p)}(q)$, and $B_{5}^{(p)}(q)$, we find similar expansions with different coefficients 
collected in Appendix \ref{app:B-funcs}.
In particular, we see that all functions $B_{k}$ are proportional to  $B_{1}$ according to 
\be
\la{4.23}
B_{2} = B_{1}, \qquad B_{3}=\frac{1}{2!}B_{1}, \qquad B_{4}=\frac{1}{3!}B_{1}, \qquad B_{5}=\frac{1}{4!}B_{1}, \quad \dots\, .
\ee
Using  the expressions in Appendix \ref{app:B-funcs}, the differential constraints (\ref{4.9}--\ref{4.13}) are readily checked to be satisfied. This proves
that the unflavored index is reproduced in the limit $u\to 1$ at all orders in $q$, up to wrapping 5.

\section{Finite $N$ corrections in   near-unflavored regime}
\la{sec:near}

We can go easily beyond the unflavored limit and give the exact $q$ dependence of the giant graviton expansion coefficients 
 in the near unflavored regime, \ie at first non-trivial order in $u-1$. From the expressions in Appendix \ref{app:B-funcs}, 
 the expansion 
of $\widehat\I^{\rm D3}_{n}(u;q)$ around $u=1$ gives the following corrections to the unflavored brane indices in  (\ref{4.4})
\bea
\la{5.1}
\widehat \I^{\rm D3}_{1} & (u; q)  = q\, \big[-(2+N)+\Delta_{1}^{(2)}(q; N)\, (u-1)^{2}+\mc O((u-1)^{3})\big], \\
\widehat \I^{\rm D3}_{2} & (u; q)  = q^{4}\, \big[\frac{1}{2}(1+N)(4+N)+\Delta_{2}^{(2)}(q; N)\, (u-1)^{2}+\mc O((u-1)^{3})\big], \\
\widehat \I^{\rm D3}_{3} & (u; q)  =q^{9}\, \big[-\frac{1}{6} (1+N)(2+N)(6+N)+\Delta_{3}^{(2)}(q; N)(u-1)^{2}+\mc O((u-1)^{3})\big], 
\eea
where $\Delta_{n}^{(2)}(q)$ admit  the exact quasimodular representations
\bea
\la{5.2}
\Delta_{1}^{(2)}(q; N) &= \frac{1}{72}\, [-12(2+N)((2+N)^{2}- \E_{2})-\E_{2}^{2}+\E_{4}], \\
\Delta_{2}^{(2)}(q; N) &= \frac{1}{72}\,[12(-2+N)(4+N)((4+N)^{2}- \E_{2})+N\,(\E_{2}^{2}-\E_{4})], \\
\Delta_{3}^{(2)}(q; N) &= \frac{1}{144}\, [-12\,(6-3N+N^{2})(6+N)((6+N)^{2}-\E_{2})-(2+N+N^{2})(\E_{2}^{2}-\E_{4})]\, .
\eea
Just to give an example, we consider the $U(2)$ index and denote $\delta u = u-1$.  We have 
\ba
\la{5.3}
& \frac{\I^{U(2)}(u; q)}{\I^{\rm KK}(u;q)} = 1-4q^{3}+9q^{8}-16q^{15}+\cdots + \big(
-10 q^3-12 q^4-24 q^5-16 q^6+24 q^8\lp
+88 q^9+48 q^{10}+144 q^{11}+36 
q^{12}+192 q^{13}
-240 q^{14}+280 q^{15}-504 q^{16}+\cdots
\big)\, \delta u^{2}+\cdots\, ,
\ea
and, setting $N=2$ in (\ref{5.1}) and (\ref{5.2}),
\bea
\la{5.4}
q^{2}\, \widehat\I^{\rm D3}_{1}(u; q) &= -4q^{3}+\big(
-10 q^3-12 q^4-24 q^5-16 q^6+24 q^8+96 q^9 
 +96 q^{10} \\
 & +240 q^{11} +260 
q^{12}+432 q^{13}+336 q^{14}+896 q^{15}+504 q^{16}+\cdots
\big)\delta u^{2}+\cdots\, ,\\
q^{2\times 2}\, \widehat\I^{\rm D3}_{2}(u; q) &= 9q^{8}+\big(
-8 q^9-48 q^{10}-96 q^{11}-224 q^{12}-240 q^{13} 
 -576 q^{14} \\
 & -448 
q^{15}-960 q^{16}+\cdots
\big)\delta u^{2}+\cdots\, ,\\
q^{3\times 2}\, \widehat\I^{\rm D3}_{2}(u; q) &= -16q^{15}+\big(
-168q^{15}-48 q^{16}+\cdots\big)\delta u^{2}+\cdots\, .
\eea
One can check that the three terms $1+q^{2}\, \widehat\I^{\rm D3}_{1}+q^{4}\, \widehat\I^{\rm D3}_{2}+q^{6}\, \widehat\I^{\rm D3}_{3}$ are enough to reproduce all terms in the r.h.s. of (\ref{5.3}).

Similar expressions may be computed at higher order in $u-1$. This requires  some additional  computational effort, but no conceptual
difficulty. For instance, for the leading wrapping correction $n=1$ we can write
\be
\widehat \I^{\rm D3}_{1}(u; q)  = q\, \big[-(2+N)+\sum_{p=2}^{\infty}\Delta_{1}^{(p)}(q; N)\, (u-1)^{p}\big]\, ,
\ee 
where the $p=2$ term was in the first line of (\ref{5.2}) and the next ones are
\ba
\Delta_{1}^{(3)}(q; N) &= -\Delta_{1}^{(2)}(q), \\
\Delta_{1}^{(4)}(q; N) &= -\frac{1}{725760}\,\bigg[-63 (14527+28800 N+18240 N^2+5600 N^3+960 N^4+96 N^5)\lp
-210 (1+144 N+24 
N^2) \,\E_{2}^2+140 (-11+6 N)\,\E_{2}^{3}+105\,\E_{2}^{4}\lp
+84 (19+192 N+60 N^2) \, \E_{4}+150\, \E_{4}^2+\E_{2}\, \bigg(60 (3155+5880 N+2016 N^2+336 N^3)\lp
-504 (-1+5 N)\,\E_{4}-240\,\E_{6}\bigg)+112 (-2+15 N)\, \E_{6}\bigg]\, , \\
\Delta_{1}^{(5)}(q; N) &= \frac{1}{362880}\, \bigg[
63 (6847+17280 N+12480 N^2+4640 N^3+960 N^4+96 N^5)\lp
+210 (-23+144 N+24 
N^2) \,\E_{2}^2-140 (-11+6 N) \,\E_{2}^3-105\, \E_{2}^4\lp
-84 (-41+192 N+60 N^2) \,\E_{4}-150 \,
\E_{4}^2
+\E_{2}\, \bigg(-60 (1139+4872 
N+2016 N^2+336 N^3)\lp
+504 (-1+5 N)\,\E_{4}+240\,\E_{6}\bigg)-112 (-2+15 N)\, \E_{6}
\bigg]\, .
\ea

\section{Giant graviton expansion of $\N=2$ $\widehat{\Gamma}(SU(N))$ theories}
\la{sec:gamma}

As a further application of our analysis, we consider in this section the 4d $\N=2$ superconformal theories theories proposed in \cite{Buican:2020moo,Kang:2021lic}. 
These models are generically non-Lagrangian and have equal conformal anomaly coefficients $a=c$. 
They are denoted $\widehat{\Gamma}(G)$, where $\Gamma$ and $G$ are ADE simply laced Lie groups, and 
are built by gauging part of the flavor symmetry of a product of the $D_{p}(G)$ superconformal theories studied in \cite{Cecotti:2012jx,Cecotti:2013lda}.
Here, we focus on the cases $\Gamma=D_{4}, E_{6}, E_{7},E_{8}$ and $G=SU(N)$, listed  in Table \ref{tab:gamma}. 
\begin{table}[H]
\be
\def\arraystretch{1.3}
\begin{array}{ccccc}
\toprule
\widehat{\Gamma}(G) & \widehat{D}_{4}(SU(2n+1)) & \widehat{E}_{6}(SU(3n\pm 1)) & \widehat{E}_{7}(SU(4n\pm 1)) & \widehat{E}_{8}(SU(6n\pm 1)) \\
\midrule
a=c &  2n(n+1) & 2n(3n\pm 2) & 6n(2n\pm 1) & 10n(3n\pm 1) \\
\bottomrule
\end{array}\notag
\ee
\caption{$\widehat{\Gamma}(SU(N))$ four-dimensional $\N=2$ superconformal 
theories with $\Gamma=D_{4}, E_{6}, E_{7}, E_{8}$ and $a=c$. The index $n$ is a positive integer. Notice that three cases 
admits the description in \cite{Xie:2016evu} as Argyres-Douglas theories engineered from M5 brane, 
$\widehat{E}_{6}(SU(2)) = (A_{2}, D_{4})$, 
$\widehat{E}_{7}(SU(3)) = (A_{3}, E_{6})$, 
$\widehat{E}_{8}(SU(5)) = (A_{5}, E_{8})$.}
\la{tab:gamma}
\end{table}
In all of these theories,  the unflavored Schur index is same as the following  
specialization of the flavored Schur index of $\N = 4$ $SU(N)$ SYM theory \cite{Buican:2020moo,Kang:2021lic}
\bea
\la{6.1}
& \I^{\widehat{\Gamma}(SU(N))}(q) = \bar\I^{SU(N)}(q^{-1}; q^{\mg}), \\
& m_{D_{4}} = 2, \quad m_{E_{6}} = 3, \quad m_{E_{7}} = 4, \quad m_{E_{8}}=6\, ,
\eea
where $\mg$ is the largest comark of the affine Dynkin
diagram $\widehat\Gamma$. 
We now show how the remarkable relation (\ref{6.1}) may be combined with the closed formulas presented for the flavored Schur index of $\N=4$ $U(N)$ SYM
in order to  derive the giant graviton expansion  of the unflavored Schur index of the $\widehat\Gamma(SU(N))$ theories.

\subsection{$\Gamma=D_{4}$}

Let us begin with the simplest case  $\Gamma=D_{4}$. We have from (\ref{6.1}), \cf (\ref{2.5}),
\be
\la{6.2}
\I^{\widehat{D}_{4}(SU(N))}(q) = \bar\I^{SU(N)}(q^{-1}; q^{2})  =\widetilde \I^{SU(N)}(q^{2})\,,
\ee
where the index in the r.h.s. is the unflavored one, \ie at  $u=1$.
Also, we recall that 
\be
\bar \I^{SU(N)}(\xi; q) = \frac{\bar\I^{U(N)}(\xi; q)}{\bar\I^{U(1)}(\xi; q)}\,.
\ee
Taking this relation in unflavored limit and using (\ref{2.12}) we get 
\be
\I^{\widehat{D}_{4}(SU(N))}(q) = \frac{(q; q^{2})_{\infty}^{2}}{(q^{2})^{2}_{\infty}}\,\widetilde\I^{U(N)}(q^{2}) = \frac{(q; q^{2})_{\infty}^{2}}{(q^{2})^{2}_{\infty}}\,\I^{U(N)}(q) \, , 
\ee
as argued in \cite{Kang:2021lic}. This case is trivial in the sense that its giant graviton expansion follows directly from the unflavored index expansion (\ref{1.8}). 
The presence of factors of $N$ in the giant graviton expansion of the $\Gamma = D_{4}$ index comes
thus from the fact that relation (\ref{6.1}) involves the unrefined index in the right hand side.

\subsection{$\Gamma=E_{6}$}

A more interesting case is $\Gamma = E_{6}$. Relation (\ref{6.1}) gives now
\be
\I^{\widehat{E}_{6}(SU(N))}(q) = \bar\I^{SU(N)}(q^{-1}; q^{3}) = \widetilde\I^{SU(N)}(q^{\frac{1}{2}}; q^{3}) =  
\frac{(q; q^{3})_{\infty}(q^{2}; q^{3})_{\infty}}{(q^{3})^{2}_{\infty}}\ \widetilde\I^{U(N)}(q^{\frac{1}{2}}, q^{3})\,.
\ee
From (\ref{1.5}), we have \footnote{By analogy with the $\N=4$ SYM case, we use the label KK for the $N=\infty$ limit of the index, 
although we don't have here a clear Kaluza-Klein interpretation
of this contribution.} 
\be
\la{6.6}
\widetilde \I^{U(\infty)}(u; q) \equiv \widetilde \I^{\rm KK}(u;q) = \frac{(q)_{\infty}}{(uq^{\frac{1}{2}})_{\infty}(u^{-1}q^{\frac{1}{2}})_{\infty}}\, ,
\ee
and in particular
\be
\widetilde \I^{U(\infty)}(q^{\frac{1}{2}}; q^{3})  =  \frac{(q^{3})_{\infty}}{(q)_{\infty}(q^{2})_{\infty}}\, .
\ee
Thus, we can write
\be
\I^{\widehat{E}_{6}(SU(N))}(q) = 
\frac{(q; q^{3})_{\infty}(q^{2}; q^{3})_{\infty}}{(q)_{\infty}(q^{2})_{\infty}(q^{3})_{\infty}}\
\frac{\widetilde\I^{U(N)}(q^{\frac{1}{2}}; q^{3})}{\widetilde \I^{U(\infty)}(q^{\frac{1}{2}};q^{3})}\, .
\ee
The ratio can be evaluated by specializing (\ref{4.1})  at $u=q^{\frac{1}{2}}$
\ba
\la{6.9}
\frac{\widetilde\I^{U(N)}(q^{\frac{1}{2}}; q^{3})}{\widetilde\I^{U(\infty)}(q^{\frac{1}{2}};q^{3})} &= 
\frac{\I^{U(N)}(q^{\frac{1}{2}}; q^{\frac{3}{2}})}{\I^{U(\infty)}(q^{\frac{1}{2}};q^{\frac{3}{2}})} =
\sum_{n=0}^{\infty}\sum_{p=0}^{n}  q^{(n+p)N+3(n-p)p}\I^{\rm D3}_{n-p}(q^{\frac{1}{2}}; q^{\frac{3}{2}}) \I^{\rm D3}_{p}(q^{-\frac{1}{2}}; q^{\frac{3}{2}})\, .
\ea
The next-to-leading giant graviton expansion is thus
\ba
\la{6.10}
\I^{\widehat{E}_{6}(SU(N))}(q) &=
\frac{(q; q^{3})_{\infty}(q^{2}; q^{3})_{\infty}}{(q)_{\infty}(q^{2})_{\infty}(q^{3})_{\infty}}
\bigg[
1+q^{N}W_{1}(q)+q^{2N}\, W_{2}(q)+\mc O(q^{3N})
\bigg]\, ,
\ea
with the explicit functions
\be
W_{1}(q) = \I_{1}^{\rm D3}(q^{-\frac{1}{2}}, q^{\frac{3}{2}}), \qquad W_{2}(q) =  \I_{1}^{\rm D3}(q^{\frac{1}{2}}; q^{\frac{3}{2}})+\I_{2}^{\rm D3}(q^{-\frac{1}{2}}; q^{\frac{3}{2}})\, .
\ee
From (\ref{3.3}), the leading term $W_{1}(q)$ is provided by the exact expression
\ba
\la{6.12}
W_{1}(q) = -q^{-\frac{3}{2}}\frac{(q^{2})^{3}_{\infty}}{
\vth(q^{3}; q^{2})} = \frac{(q^{2})_{\infty}^{2}}{(q^{-1}; q^{2})_{\infty}(q^{3};q^{2})_{\infty}} = -q\, \frac{(q^{2})_{\infty}^{2}}{(q; q^{2})_{\infty}^{2}}\, ,
\ea
and is remarkably $N$ independent. Notice that in terms of Jacobi elliptic theta functions, \cf (\ref{jac}), we can also write $W_{1}$ as 
\be
W_{1}(q) = -\frac{1}{4}\, q^{\frac{3}{4}}\, \vth_{2}(\sqrt{q})^{2}\, .
\ee

The next-to-leading term $W_{2}(q)$ is the sum of two contributions that are separately singular, suggesting some extra factor of $N$ appearing at double wrapping order. 
Again, this is not surprising since 
the key relation (\ref{6.1}) involves the flavored Schur index evaluated with a special algebraic relation between fugacities, as it happened in the $\Gamma=D_{4}$ case. Singularities cancel in the sum and 
to get the finite correction we have to regularize the 
flavor fugacity by scaling $q^{\frac{1}{2}}\to \eta q^{\frac{1}{2}}$ in the $u$ argument in (\ref{6.9}) and taking $\eta\to 1$ in the end. This gives 
\be
\la{6.14}
W_{2}(q) = \lim_{\eta\to 1}W_{2}(\eta; q), \qquad W_{2}(\eta; q) = \eta^{N}\I_{1}^{\rm D3}(\eta\, q^{\frac{1}{2}}; q^{\frac{3}{2}})+\eta^{-2N}\I_{2}^{\rm D3}(\eta^{-1}q^{-\frac{1}{2}}; q^{\frac{3}{2}}) 
\ee
The derivation of the limit $W_{2}(q)$ is a little involved and is discussed in Appendix \ref{app:W2}. The result  is 
\be
\la{6.15}
W_{2}(q) = \bigg(N+\frac{23}{8}\bigg)\, q^{3}-\frac{q}{4}R_{2}(q^{-1}; q^{2})+\frac{1}{6}R_{3}(q^{-1}; q^{2})\, , 
\ee
where we remind that the functions $R_{k}$ are related to derivatives of $\log\theta$ according to (\ref{A.9}). A more explicit expression is, \cf definitions in (\ref{jac}),
\be
W_{2}(q) = q^{3}\, \bigg[N+3+\frac{1}{8}\frac{\partial^{2}_{z}\vth_{4}(0; q)}{\vth_{4}(q)}\bigg]\, .
\ee
The expansion of $W_{2}(q)$ in powers of $q$ reads 
\be
W_{2}(q) = (N+3)q^{3}+q^{4}+2q^{5}+4q^{6}+4q^{7}+6q^{8}+8q^{9}+\cdots\, ,
\ee
with a peculiar $N$ dependence in the $q^{3}$ term only.
In summary, the explicit giant graviton expansion of the $\widehat{E}_{6}(SU(N))$ Schur index is 
\ba
\la{618}
\I^{\widehat{E}_{6}(SU(N))}(q) &=
\frac{(q; q^{3})_{\infty}(q^{2}; q^{3})_{\infty}}{(q)_{\infty}( q^{2})_{\infty}(q^{3})_{\infty}}
\bigg[ 
1+q^{N}(-q-2q^{2}-q^{3}-2q^{4}-2q^{5}-3q^{7}-2q^{8}+\mc O(q^{10}) \cdots)\lp
+q^{2N}\bigg((N+3)q^{3}+q^{4}+2q^{5}+4q^{6}+4q^{7}+6q^{8}+8q^{9}+\mc O(q^{10})\bigg) +\mc O(q^{3N+7})
\bigg]\,.
\ea
Notice that the first omitted terms are not just $\sim q^{3N}$ since one finds an leading term $-5q^{7}$ in 
the expression  for $W_{3}$, which is the first contribution omitted in (\ref{6.10}).

\paragraph{Comparison with available data}

As a check, the final expansion (\ref{618}) may be compared with the 
explicit results computed in \cite{Kang:2021lic}. They read \footnote{
The $\widehat{E}_{6}(SU(2))$ 
case is known in closed form as $\I^{\widehat{E}_{6}(SU(2))}(q)=\sum_{n=1}^{\infty}\frac{q^{n-1}}{1+q^{n}+q^{2n}}$ \cite{Hatsuda:2022xdv}.}
\bea
\I^{\widehat{E}_{6}(SU(2))} &= 1+q^{2}+q^{3}+2q^{6}+q^{8}+q^{11}+2q^{12}+q^{15}+2q^{18}+\cdots, \\
\I^{\widehat{E}_{6}(SU(4))} &= 1+q^{2}+2q^{3}+2q^{4}+q^{5}+6q^{6}+2q^{7}+4q^{8}+7q^{9}+7q^{10}+4q^{11}+\cdots, \\
\I^{\widehat{E}_{6}(SU(5))} &= 1+q^{2}+2q^{3}+2q^{4}+2q^{5}+7q^{6}+2q^{7}+8q^{8}+10q^{9}+8q^{10}+\cdots\, .
\eea
Dividing by the Kaluza-Klein factor in (\ref{6.10}), we find 
\bea
\la{6.19}
\I^{\widehat{E}_{6}(SU(2))}/\I^{\rm KK} &= 1-q^{3}-2q^{4}-q^{5}-2q^{6}+3q^{7}+q^{8}-q^{9}+2q^{10} +\cdots, \\
\I^{\widehat{E}_{6}(SU(4))}/\I^{\rm KK}  &= 1-q^{5}-2q^{6}-q^{7}-2q^{8}-2q^{9}+4q^{11}+\cdots, \\
\I^{\widehat{E}_{6}(SU(5))}/\I^{\rm KK}  &= 1-q^{6}-2q^{7}-q^{8}-2q^{9}-2q^{10}+\cdots\, .
\eea
To reproduce these series, it is enough to use the two terms we computed, \ie 
\be
1+q^{N}W_{1}+q^{2N}W_{2}+\cdots\, .
\ee
For instance for $N=2$ one has, up to terms of order $q^{10}$ included
\ba
& 1+q^{2}(-q-2q^{2}-q^{3}-2q^{4}-2q^{5}-3q^{7}-2q^{8}+\cdots)
+q^{4}\big(\bm{5}\,q^{3}+\bm{1}\,q^{4}+\bm{2}\,q^{5}+\bm{4}\,q^{6}+\cdots\bigg) +\mc O(q^{11}) \lp
= 1-q^{3}-2q^{4}-q^{5}-2q^{6}+(-2+\bm{5})\, q^{7}+(0+\bm{1})\,q^{8}+(-3+\bm{2})q^{9}+(-2+\bm{4})\,q^{10}+\mc O(q^{11})\, ,
\ea
in agreement with the first line in (\ref{6.19}). The other cases can be checked similarly.

\subsection{$\Gamma=E_{7},E_{8}$}

For a generic $\mg$, we can repeat the same steps. Let us discuss in particular the leading wrapping correction. 
Relation (\ref{6.1}) gives
\ba
\I^{\widehat{\Gamma}(SU(N))}(q) &= \bar\I^{SU(N)}(q^{-1}; q^{\mg}) = \widetilde\I^{SU(N)}(q^{\frac{\mg-2}{2}}; q^{\mg}) \lp
=  \frac{(q; q^{\mg})_{\infty}(q^{\mg-1}; q^{\mg})_{\infty}}{(q^{\mg})^{2}_{\infty}}\
 \widetilde\I^{U(N)}(q^{\frac{\mg-2}{2}}, q^{\mg})\,.
\ea
From (\ref{6.6}) we have the Kaluza-Klein factor
\be
\widetilde \I^{U(\infty)}(q^{\frac{\mg-2}{2}}; q^{\mg})= \frac{(q^{\mg})_{\infty}}{(q^{\mg-1})_{\infty}(q)_{\infty}}\, ,
\ee
and thus
\be
\I^{\widehat{\Gamma}(SU(N))}(q) =
\frac{(q; q^{\mg})_{\infty}(q^{\mg-1}; q^{\mg})_{\infty}}{(q)_{\infty}(q^{\mg-1})_{\infty}(q^{\mg})_{\infty}}\
\frac{\widetilde\I^{U(N)}(q^{\frac{\mg-2}{2}}; q^{\mg})}{\widetilde\I^{U(\infty)}(q^{\frac{\mg-2}{2}};q^{\mg})}\, .
\ee
The ratio can be evaluated by specializing (\ref{4.1})  at $u=q^{\frac{\mg-2}{2}}$
\ba
\frac{\widetilde\I^{U(N)}(q^{\frac{\mg-2}{2}}; q^{\mg})}{\widetilde\I^{U(\infty)}(q^{\frac{\mg-2}{2}};q^{\mg})} &= 
\frac{\I^{U(N)}(q^{\frac{\mg-2}{2}}; q^{\frac{\mg}{2}})}{\I^{U(\infty)}(q^{\frac{\mg-2}{2}};q^{\frac{\mg}{2}})} =
\bigg[
1+q^{N}\, W_{1}^{\Gamma}(q)+\mc O(q^{2N})\bigg]\,,
\ea
where the  leading order wrapping correction reads
\be
W_{1}^{\Gamma}(q) = \I_{1}^{\rm D3}(q^{-\frac{\mg-2}{2}}, q^{\frac{\mg}{2}}) = -q\,\frac{(q^{\mg-1})_{\infty}^{2}}{(q;q^{\mg-1})_{\infty}(q^{\mg-2},q^{\mg-1})_{\infty}}\,.
\ee
For $\mg=3$ it reproduces (\ref{6.12}). In the $E_{7},E_{8}$ cases, \ie for $\mg=4,6$, 
its series expansion  reads
\bea
W_{1}^{E_{7}}(q) &= -q-q^2-2 q^3-2 q^5-q^6-2 q^7-q^9+\mc O(q^{10}), \\
W_{1}^{E_{8}}(q) &= -q-q^2-q^3-q^4-2 q^5-q^7-q^8-2 q^9+\mc O(q^{10})\,.
\eea

\section*{Acknowledgements}

We thank Arkady Tseytlin, Matthew Buican, Yosuke Imamura, and Ji Hoon Lee for useful discussions related to various aspects of this work. 
Financial support from the INFN grant GAST is acknowledged. ACB would like to thank the Isaac Newton Institute for Mathematical Sciences, Cambridge, for support and hospitality during the programme \textit{Black holes: bridges between number theory and holographic quantum information}, where work on this paper was undertaken. This work was supported by EPSRC grant EP/R014604/1.

\appendix
\section{Special functions}
\la{app:special}

We collect in this appendix the definition of  special functions appearing in the text and some useful identities.

\paragraph{Jacobi elliptic theta functions}

\bea
\la{jac}
\vth_{1}(z; q) &= 2\,q^{\frac{1}{4}}\,\sum_{n=0}^{\infty}(-1)^{n}\, \sin[(2n+1)\,z]\, q^{n\,(n+1)}\, , \\
\vth_{2}(z; q) &= 2\,q^{\frac{1}{4}}\,\sum_{n=0}^{\infty}\cos[(2n+1)\,z]\, q^{n\,(n+1)}\, , \\
\vth_{3}(z; q) &= 1+2\,q^{\frac{1}{4}}\,\sum_{n=1}^{\infty} \cos(2\,n\,z)\, q^{n^{2}}\, , \\
\vth_{4}(z; q) &= 1+2\,\sum_{n=1}^{\infty} (-1)^{n}\,\cos(2\, n\, z)\, q^{n^{2}}\, . 
\eea
Also, $\vth_{n}(q) \equiv \vth_{n}(0; q)$, $n=2,3,4$.

\paragraph{Dedekind $\eta$ function}
\be
\la{A.1}
\eta(\tau) = q^{\frac{1}{12}}\prod_{k=1}^{\infty}(1-q^{2k}), \qquad q=e^{i\pi \tau}\,.
\ee

\paragraph{$q$-Pochhammer symbol}

\ba
\la{A.4}
(a; q)_{\infty}&= \prod_{k=0}^{\infty}(1-a\,q^{k})\,, \qquad (a^{\pm}; q)_{\infty} = (a; q)_{\infty}(a^{-1};  q)_{\infty}\, , \\
(q)_{\infty} &\equiv (q; q)_{\infty} = \prod_{k=1}^{\infty}(1-q^{k})\, .
\ea
Notice that we can write the Dedekind function in (\ref{A.1}) as 
\be
\la{A.4}
\eta(\tau) = q^{\frac{1}{12}}(q^{2})_{\infty}\,.
\ee
An elementary but useful  relation is 
\be
\la{A.5}
(aq; q)_{\infty} = \frac{1}{1-a}(a; q)_{\infty}\, .
\ee

\paragraph{$q$-theta function}

The $q$-theta function is defined as 
\be
\la{A.6}
\vth(x,q) = -x^{-\frac{1}{2}}(q)_{\infty}(x; q)_{\infty}(qx^{-1}; q)_{\infty}\, ,  
\ee
with 
\ba
\la{A.7}
\vth(x;q) = -\vth(x^{-1};q), \qquad 
\vth'(x;q) = \frac{1}{x^{2}}\vth'(x^{-1};q), \qquad 
\vth''(x;q) = -\frac{2}{x^{3}}\vth'(x^{-1};q)-\frac{1}{x^{4}}\vth''(x^{-1};q)\,. 
\ea
It obeys the very useful relation
\be
\la{A.8}
\vth(q^{m}x; q) = (-1)^{m}q^{-\frac{m^{2}}{2}}x^{-m}\vth(x;q)\,.
\ee
Introducing the ratios
\be
\la{A.9}
R_{k}(x;q) = \frac{\vth^{(k)}(x;q)}{\vth(x;q)}, \qquad \vth^{(k)}(x;q) = \partial_{x}^{k}\vth(x; q)\, ,
\ee
we obtain from (\ref{A.8}) the relations
\ba
\la{A.10}
R_{k}(q^{m}x; q) &= \sum_{p=0}^{k}(-1)^{p}\binom{k}{p}\frac{(m)_{p}}{x^{p}}\, R_{k-p}(x;q)\,.
\ea
%

\paragraph{Eisenstein series}

The classical Eisenstein series are defined as 
\be
\la{A.11}
\E_{2m}(q) = 1-\frac{4m}{B_{2m}}\sum_{n=1}\frac{n^{2m-1}q^{n}}{1-q^{n}}\, ,
\ee
where $B_{2m}$ are Bernoulli numbers. We will need the differential equations
\ba
\la{A.12}
q\frac{d}{dq}\E_{2}(q) &= \frac{1}{12}(\E_{2}^{2}-\E_{4}), \\
\la{A.13}
q \frac{d}{dq}\E_{4}(q) &= \frac{1}{3}(\E_{2}\E_{4}-\E_{6}), \\
q \frac{d}{dq}\E_{6}(q) &= \frac{1}{2}(\E_{2}\E_{6}-\E_{4}^{2})\,.
\ea

\paragraph{Twisted Weierstrass functions}
\bea
\la{A.14}
\wei{1}{x}{1}(\nu,\tau) &= \frac{(q)_{\infty}^{3}\vth(x^{-1}w; q)}{\vth(x^{-1}; q)\vth(w;q)}, \qquad w=e^{2\pi i \nu}, \\
\wei{k}{x}{q}(\nu,\tau) &= \frac{(-1)^{k-1}}{(k-1)!}\frac{(q)_{\infty}^{3}}{\vth(x^{-1}; q)} \left(w\frac{\partial}{\partial w}\right)^{k-1}\frac{\vth(x^{-1}w; q)}{\vth(w;q)}\, .
\eea

%

\section{Technical details}

\subsection{Generalized Lambert series and the sums $S_{3}, S_{4}$}
\la{app:eis}

Let us consider the generalized Lambert series
\be
L_{a,b}(q)=\sum_{k=1}^{\infty} \frac{k^{b}q^{k}}{(1-q^{k})^{a}}\,.
\ee
The cases $L_{1,b}$ with odd $b$ are directly related to Eisenstein series (\ref{A.11}), but other cases are not. For instance the  sum
$L_{1,0}$ is the generating function of the number of positive divisors and can be written in terms of the $q$-polygamma function, but this is actually the definition of that function.
We will adopt  the following abbreviated notation for the sums $L_{1,b}$ with  even index $b$
\be
L_{2m}(q) \equiv L_{1,2m}(q) = \sum_{k=1}^{\infty}\frac{k^{2m}q^{k}}{1-q^{k}}\, , \qquad m=0,1,2,\dots\, .
\ee
We now show how to write $L_{a,b}$ in terms of Eisenstein series, sums of type $L_{2m}$, and their derivatives. To this aim, 
we start by remarking that 
a large set of  identities among  $L_{a,b}$ with different indices comes after summing over $\ell$ in the trivial  identity 
\be
\la{C.3}
\sum_{k,\ell=1}^{\infty}(k-\ell) \times (\text{symmetric polynomial in $k,\ell$})\,q^{k\ell} = 0 .
\ee
For example, 
\be
\sum_{k,\ell=1}^{\infty}(k-\ell)q^{k\ell}=0\quad \to\quad\sum_{k=1}^{\infty}\bigg[\frac{k}{1-q^{k}}-\frac{1}{(1-q^{k})^{2}}\bigg]\, q^{k}= 0\,,
\ee
gives the relation
\be
L_{1,1}=L_{2,0}\,.
\ee
Another useful set of relations is obtained by differentiating with respect to $q$, for instance
\be
\la{C.6}
q\frac{d}{dq}\E_{2}=-24q\frac{d}{dq}\sum_{k=1}^{\infty}\frac{k q^{k}}{1-q^{k}} = -24\sum_{k=1}^{\infty}\frac{k^{2}q^{k}}{(1-q^{k})^{2}} = -24\, L_{2,2}\,.
\ee
Considering all instances of (\ref{C.3}) with a generic symmetric polynomial of degree 4 and adding to (\ref{C.6}) the analogous relation for $q^{2}\E_{2}''$, one gets the following
relations for $L_{a,b}$. With first index equal to 2:
\begin{alignat}{2}
L_{2,0} &= \frac{1}{24}(1-\E_{2}), \qquad && L_{2,3} = q L_{2}', \notag  \\
L_{2,1} &= q\, L_{0}', && L_{2,4} = \frac{1}{240}q\E_{4}'\, \\
L_{2,2} &= -\frac{1}{24}q \E_{2}'\, . \notag 
\end{alignat}
With first index 3:
\begin{alignat}{2}
L_{3,0} &= \frac{1}{48}(1-\E_{2})+\frac{1}{2}L_{2}, \qquad && L_{3,2} = -\frac{1}{48}q\E_{2}'+\frac{1}{2}qL_{0}'+\frac{1}{2}q^{2}L_{0}'',\notag \\
L_{3,1} &= -\frac{1}{48}q\E_{2}'+\frac{1}{2}qL_{0}', && L_{3,3} = -\frac{1}{48}q\E_{2}'-\frac{1}{48}q^{2}\E_{2}''+\frac{1}{2}qL_{2}'\, .
\end{alignat}
With first index 4, 5, 6
\ba
L_{4,0} &= \frac{19}{1440}-\frac{1}{72}\E_{2}+\frac{1}{1440}\E_{4}+\frac{1}{2}L_{2}, \notag \\
L_{4,1} &= -\frac{1}{48}q\E_{2}'+\frac{1}{3}qL_{0}'+\frac{1}{6}qL_{2}', \notag \\
L_{4,2} &= -\frac{1}{48}\E_{2}'-\frac{1}{144}q^{2}\E_{2}''+\frac{1}{2}qL_{0}'+\frac{1}{2}q^{2}L_{0}'', \\
L_{5,0} &= \frac{3}{320}-\frac{1}{96}\E_{2}+\frac{1}{960}\E_{4}+\frac{11}{24}L_{2}+\frac{1}{24}L_{4}, \notag \\
L_{5,1} &= -\frac{11}{576}q \E_{2}'+\frac{1}{5760}q\E_{4}'+\frac{1}{4}qL_{0}'+\frac{1}{4}qL_{2}', \notag \\
L_{6,0} &= \frac{863}{120960}-\frac{1}{120}\E_{2}+\frac{7}{5760}\E_{4}-\frac{1}{60480}\E_{6}+\frac{5}{12}L_{2}+\frac{1}{12}L_{4}\, .\notag 
\ea

\paragraph{The sum $S_{3}(q)$}

Let us now consider the sum  in (\ref{4.17}). Factoring $q^{k}$ and splitting $q^{k}= 1-(1-q^{k})$ in the rest, we get 
\be
S_{3}(q) = \sum_{k=1}^{\infty}  \frac{4(1+k+(k-1) q^{k})q^{k}}{(1-q^{k})^{3}} = 4L_{2,0}-4L_{2,1}+8L_{3,1}.
\ee
Using the previous relations, we obtain 
\be
S_{3}(q) =  \frac{1}{6}(1-\E_{2}-q \E_{2}')\, ,
\ee
which is same as (\ref{4.20}) using the  differential equation (\ref{A.12}).

\paragraph{The sum $S_{4}(q)$} The procedure for this sum is similar. First, we write 
\ba
S_{4}(q) &= \sum_{k=1}^{\infty} \frac{[5+6k+2k^{2}-2(1-4k^{2})q^{k}+(5-6k+2k^{2})q^{2k}]q^{k}}{(1-q^{k})^{4}}  \lp
= 5 L_{2,0} - 6 L_{2,1} + 2 L_{2,2} - 8 L_{3,0} + 12 L_{3,1} - 
 12 L_{3,2} + 8 L_{4,0} + 12 L_{4,2}\, .
\ea
Then, using the previous relations, we get 
\be
S_{4}(q) =  \frac{53}{360}-\frac{11}{72}\E_{2}+\frac{1}{180}\E_{4}
-\frac{1}{3}q\E_{2}'-\frac{1}{12}q^{2}\E_{2}''\, .
\ee
This is same as (\ref{4.21}) after using (\ref{A.12}) and (\ref{A.13})
that give
\ba
q^{2}\E_{2}'' &=\bigg(q\frac{d}{dq}\bigg)^{2}\E_{2}-q\frac{d}{dq}\E_{2} = 
\frac{1}{72}\E_{2}^{3}-\frac{1}{24}\E_{2}\E_{4}+\frac{1}{36}\E_{6}-\frac{1}{12}\E_{2}^{2}+\frac{1}{12}\E_{4}.
\ea

\subsection{Differential equation for $B_{1}(u;q)$}
\la{C.2}

Let us start from the single wrapping brane index
\ba
\I^{\rm D3}_{1}(u; q) &=  -qu^{2}\frac{({\frac{q}{u}})_{\infty}^{3}}{\vth(u^{2}; \frac{q}{u})}\,,
\ea
and define
\be
f(x; q) = \I^{\rm D3}_{1}(\sqrt{x}; \sqrt{x}\,q) = -q x^{3/2}\frac{(q)_{\infty}^{3}}{\vth(x;q)}\, .
\ee
From the differential equation for the $\vth$ function
\be
\la{C.17}
q\frac{\partial}{\partial q}\vth(x;q)-\frac{1}{2}\bigg(x\frac{\partial}{\partial x}\bigg)^{2}\vth(x; q)+\frac{1}{8}\vth(x;q)=0\,,
\ee
and the well known derivative 
\be
q\frac{d}{dq}\log(q)_{\infty} = \frac{1}{24}(\E_{2}-1), 
\ee
we obtain the non-linear differential equation for $f(x; q)$
\be
\la{B19}
\frac{9}{8}+\frac{1}{8}(\E_{2}-1)-\bigg(q\partial_{q}-\frac{1}{2}x\partial_{x}-\frac{1}{2}x^{2}\partial_{x}^{2}\bigg)\log f
-\frac{1}{2}x^{2}\bigg(\frac{3}{2x}-\partial_{x}\log f\bigg)^{2} = 0\, .
\ee
From (\ref{3.17}), $f(x; q)$ has the structure
\be
f(x,q) = \frac{x^{2}}{1-x}\,q+x(1-x)q^{2}B_{1}(\sqrt x; q \sqrt x)\, ,
\ee
and therefore, expanding around $x=1$, we get 
\ba
\log f(x,q) &= \log \frac{q}{1-x}+2 (x-1)+[-1+q\,B_{1}(q)]\,
(x-1)^2\lp
+\frac{1}{6} (4-6 q \,B_{1}(q)+3 q^2\,B_{1}'(q)+3 q B_{1}^{(1)}(q)]\, (x-1)^3+\mc O((x-1)^4)\, ,
\ea
where notation is as in (\ref{411}).
Plugging this expansion in (\ref{B19}) gives
\ba
0 = \frac{1}{8} (-1+24 q\, B_{1}(q)+\E_{2})+3 q [q B_{1}'(q)+B_{1}^{(1)}(q)]\, (x-1)+\mc O((x-1)^2\, .
\ea
The leading order gives
\be
\la{C.21}
B_{1}(q)\equiv B_{1}(1; q) = \frac{1}{24}\big(1-\E_{2}(q)\big).
\ee
At next orders in $x-1$, using recursively results like (\ref{C.21}) and its derivatives with respect to $q$, we get 
the following expressions for $\partial_{u}^{p}B_{1}(1; q)$ in terms of $\E_{2}(q)$ and its derivatives
\bea
q\partial_{u}B_{1}(1; q) &= \frac{1}{24}(1-\E_{2})+\frac{1}{24}\E_{2}', \\
q\partial_{u}^{2}B_{1}(1; q) &= -\frac{17}{720}+\frac{1}{72}\E_{2}+\frac{7}{720}\E_{2}^{2}-\frac{1}{30}q\E_{2}'-\frac{1}{24}q^{2}\E_{2}'', \\
q\partial_{u}^{3}B_{1}(1; q) &= \frac{7}{120} q \E_2'-\frac{7}{120} q \E_2 \E_2'+\frac{9}{40} q^2 \E_2''+\frac{1}{24} q^3 \E_2^{(3)}, \\
q\partial_{u}^{4}B_{1}(1; q) &=\frac{377}{5040}-\frac{\E_2}{30}-\frac{7 
\E_2^2}{240}-\frac{31 \E_2^3}{2520}-\frac{22}{105} q 
\E_2'+\frac{71}{210} q \E_2 \E_2'+\frac{7}{60} 
q^2 \E_2'^2-\frac{181}{140} q^2 \E_2'' \\
& +\frac{7}{60} 
q^2 \E_2 \E_2''-\frac{8}{15} q^3 \E_2^{(3)}-\frac{1}{24} q^4 \E_2^{(4)}\, ,
\eea
and so on. These results are in agreement with (\ref{4.22}), using the differential equations (\ref{A.12}) and (\ref{A.13}). This way, it is straightforward to 
compute any function $B_{1}^{(p)}(q)$ at arbitrarily large $p$.

\subsection{Quasimodular expressions for the functions $B_{n}^{(p)}(q)$}
\la{app:B-funcs}

The functions $B_{n}^{(p)}(q)$ for $n=3,4,5$ and $p=1,2,3$ have a quasimodular structure completely similar to that in (\ref{4.22}). The explicit expressions are 

\paragraph{Functions $B_{2}^{(p)}$}

\bea
q\, B_{2}(q) &= \frac{1}{24}(1-\E_{2}),\\
q\,B_{2}^{(1)}(q) &= \frac{1}{12}(1-\E_{2})+\frac{1}{288}(\E_{2}^{2}-\E_{4}), \\
q\,B_{2}^{(2)}(q)  &= -\frac{43}{30}+\frac{17}{12}\E_{2}+\frac{1}{96}\E_{2}^{2}+\frac{1}{160}\E_{4}-\frac{1}{1728}\E_{2}^{3}+\frac{1}{576}\E_{2}\E_{4}-\frac{1}{864}\E_{6},\\
q\,B_{2}^{(3)}(q)  &= -\frac{443}{20}+\frac{45}{2}\E_{2}-\frac{125}{144}\E_{2}^{2}+\frac{373}{720}\E_{4}-\frac{1}{576}\E_{2}^{3}-\frac{11}{960}\E_{2}\E_{4}
+\frac{19}{1440}\E_{6}+\frac{1}{6912}\E_{2}^{4}\\
& +\frac{1}{864}\E_{2}\E_{6}-\frac{1}{2304}\E_{4}^{2}-\frac{1}{1152}\E_{2}^{2}\E_{4}\, .
\eea

\paragraph{Functions $B_{3}^{(p)}$}

\bea
q\, B_{3}(q) &= \frac{1}{48}(1-\E_{2}),\\
q\,B_{3}^{(1)}(q) &= \frac{1}{12}(1-\E_{2})+\frac{1}{576}(\E_{2}^{2}-\E_{4}), \\
q\,B_{3}^{(2)}(q)  &= -\frac{1597}{1440}+\frac{79}{72}\E_{2}+\frac{5}{576}\E_{2}^{2}+\frac{1}{320}\E_{4}-\frac{1}{3456}\E_{2}^{3}
+\frac{1}{1152}\E_{2}\E_{4}-\frac{1}{1728}\E_{6},\\
q\,B_{3}^{(3)}(q)  &= -\frac{303}{160}+\frac{9}{4}\E_{2}-\frac{73}{96}\E_{2}^{2}+\frac{97}{240}\E_{4}-\frac{1}{1152}\E_{2}^{3}-\frac{53}{5760}\E_{2}\E_{4}
+\frac{29}{2880}\E_{6}\\
& +\frac{1}{13824}\E_{2}^{4}+\frac{1}{1728}\E_{2}\E_{6}-\frac{1}{4608}\E_{4}^{2}-\frac{1}{2304}\E_{2}^{2}\E_{4}\, .
\eea

\paragraph{Functions $B_{4}^{(p)}$}

\bea
q\, B_{4}(q) &= \frac{1}{144}(1-\E_{2}),\\
q\,B_{4}^{(1)}(q) &= \frac{7}{144}(1-\E_{2})+\frac{1}{1728}(\E_{2}^{2}-\E_{4}), \\
q\,B_{4}^{(2)}(q)  &= -\frac{343}{1080}+\frac{5}{16}\E_{2}+\frac{1}{192}\E_{2}^{2}-\frac{1}{8640}\E_{4}
-\frac{1}{10368}\E_{2}^{3}+\frac{1}{3456}\E_{2}\E_{4}-\frac{1}{5184}\E_{6}, \\
q\,B_{4}^{(3)}(q) &= \frac{1951}{240}-\frac{191}{24}\E_{2}-\frac{275}{864}\E_{2}^{2}+\frac{637}{4320}\E_{4}-\frac{1}{1728}\E_{2}^{3}
-\frac{29}{8640}\E_{2}\E_{4}\\
& +\frac{17}{4320}\E_{6}+\frac{1}{41472}\E_{2}^{4}+\frac{1}{5184}\E_{2}\E_{6}-\frac{1}{13824}\E_{4}^{2}-\frac{1}{6912}\E_{2}^{2}\E_{4}\,,
\eea

\paragraph{Functions $B_{5}^{(p)}$}

\bea
q\, B_{5}(q) &= \frac{1}{576}(1-\E_{2}),\\
q\,B_{5}^{(1)}(q) &= \frac{11}{576}(1-\E_{2})+\frac{1}{6912}(\E_{2}^{2}-\E_{4}), \\
q\,B_{5}^{(2)}(q)  &= \frac{163}{17280}-\frac{19}{1728}\E_{2}+\frac{5}{2304}\E_{2}^{2}-\frac{7}{11520}\E_{4}
-\frac{1}{41472}\E_{2}^{3}+\frac{1}{13824}\E_{2}\E_{4}-\frac{1}{20736}\E_{6}, \\
q\,B_{5}^{(3)}(q) &= \frac{17447}{2880}-\frac{865}{144}\E_{2}-\frac{23}{288}\E_{2}^{2}+\frac{83}{2880}\E_{4}
-\frac{1}{3456}\E_{2}^{3}-\frac{1}{1440}\E_{2}\E_{4}+\frac{17}{17280}\E_{6}\\
&+\frac{1}{165888}\E_{2}^{4}+\frac{1}{20736}\E_{2}\E_{6}
-\frac{1}{55296}\E_{4}^{2}-\frac{1}{27648}\E_{2}^{2}\E_{4}\,.
\eea
As an alternative basis, all functions $B_{k}^{(1)}$ can be expressed in terms of $B_{1}$ and $B_{1}^{(1)}$ according to 
\bea
\la{C.27}
B_{2}^{(1)} &= B_{1}^{(1)}+B_{1}, \qquad B_{3}^{(1)} = \frac{1}{2}B_{1}^{(1)}+\frac{3}{2}B_{1}, \\ 
B_{4}^{(1)} &= \frac{1}{6}B_{1}^{(1)}+B_{1}, \qquad
B_{5}^{(1)} = \frac{1}{24}B_{1}^{(1)}+\frac{5}{12}B_{1}\, .
\eea
Similarly, $B_{k}^{(2)}$ can be given in terms of $B_{1}, B_{1}^{(1)}, B_{1}^{(2)}$
with the addition of a term proportional to  $q B_{1}^{2}$ 
\bea
\la{C.28}
B_{2}^{(2)} &= B_{1}^{(2)}-2B_{1}^{(1)}-32 B_{1}+4q B_{1}^{2}, \\
B_{3}^{(2)} &= \frac{1}{2}B_{1}^{(2)}-B_{1}^{(1)}-\frac{51}{2} B_{1}+4q B_{1}^{2}, \\
B_{4}^{(2)} &= \frac{1}{6}B_{1}^{(2)}-\frac{137}{18} B_{1}+2q B_{1}^{2}, \\
B_{5}^{(2)} &= \frac{1}{24}B_{1}^{(2)}+\frac{1}{6}B_{1}^{(1)}+\frac{1}{18} B_{1}+\frac{2}{3}q B_{1}^{2}\,.
\eea
Third derivatives require a further contribution proportional to $qB_{1}B_{1}^{(1)}$
and are given by 
\bea
\la{C.29}
B_{2}^{(3)} &= B_{1}^{(3)}-9B_{1}^{(2)}-\frac{732}{5}B_{1}^{(1)}-384 B_{1}-\frac{876}{5}q B_{1}^{2}+24 q B_{1}B_{1}^{(1)}, \\
B_{3}^{(3)} &= \frac{1}{2}B_{1}^{(3)}-\frac{15}{2}B_{1}^{(2)}-\frac{1149}{10}B_{1}^{(1)}+ 72B_{1}-\frac{936}{5}q B_{1}^{2}+24 q B_{1}B_{1}^{(1)}, \\
B_{4}^{(3)} &= \frac{1}{6}B_{1}^{(3)}-3B_{1}^{(2)}-\frac{1261}{30}B_{1}^{(1)}+\frac{1433}{6} B_{1}-\frac{468}{5}q B_{1}^{2}+12 q B_{1}B_{1}^{(1)}, \\
B_{5}^{(3)} &= \frac{1}{24}B_{1}^{(3)}-\frac{3}{4}B_{1}^{(2)}-\frac{247}{30}B_{1}^{(1)}+\frac{617}{4} B_{1}-\frac{146}{5}q B_{1}^{2}+4 q B_{1}B_{1}^{(1)}\,.
\eea

\subsection{Evaluation of $W_{2}(q)$}
\la{app:W2}

In this Appendix, we compute the function $W_{2}(q)$ from the limit  (\ref{6.14}), \ie 
\be
\la{C.30}
W_{2}(q) = \lim_{\eta\to 1}[ \eta^{N}\I_{1}^{\rm D3}(\eta\, q^{\frac{1}{2}}; q^{\frac{3}{2}})+\eta^{-2N}\I_{2}^{\rm D3}(\eta^{-1}q^{-\frac{1}{2}}; q^{\frac{3}{2}}) ]\, .
\ee
The first term in (\ref{C.30}) can be written using (\ref{A.5}) in the form 
\ba
\I_{1}^{\rm D3}(\eta\, q^{\frac{1}{2}}; q^{\frac{3}{2}}) = -\frac{q^{3}\eta^{6}}{(1-\eta q^{2})(1-\eta^{2}q)(1-\eta^{3})}\frac{(\frac{q}{\eta})^{2}_{\infty}}
{(q^{3};\frac{q}{\eta})_{\infty}(\frac{q}{\eta^{4}};\frac{q}{\eta})_{\infty}}\, .
\ea
We now observe that 
\ba
& \log\frac{(\frac{q}{\eta})^{2}_{\infty}}{(q^{3};\frac{q}{\eta})_{\infty}(\frac{q}{\eta^{4}};\frac{q}{\eta})_{\infty}} = \sum_{n=0}^{\infty}\bigg[\log\frac{1-q^{n+1}}{1-q^{n+3}}+\bigg(
\frac{(n-2)q^{n+1}}{1-q^{n+1}}-\frac{nq^{n+3}}{1-q^{n+3}}\bigg)(\eta-1)+\mc O((\eta-1)^{2})\bigg] \lp
= \log[(1-q)(1-q^{2})]-\bigg(\frac{2q}{1-q}+\frac{q^{2}}{1-q^{2}}\bigg)(\eta-1)+\mc O((\eta-1)^{2})\bigg], 
\ea
and therefore
\ba
& \frac{(\frac{q}{\eta})^{2}_{\infty}}{(q^{3};\frac{q}{\eta})_{\infty}(\frac{q}{\eta^{4}};\frac{q}{\eta})_{\infty}} 
= (1-q)(1-q^{2})-q(1-q)(2+3q)(\eta-1)+\mc O((\eta-1)^{2})\, .
\ea
Remarkably, the term linear in $\eta-1$  is a simple polynomial in $q$, a  property thet does not hold for the next corrections quadratic in $\eta-1$.
Still, this is enough to give the following expansion of the first term in (\ref{C.30})  around $\eta=1$
\ba
\eta^{N}\, \I_{1}^{\rm D3}(\eta\, q^{\frac{1}{2}}; q^{\frac{3}{2}}) = \frac{q^{3}}{3(\eta-1)}+\frac{5+N}{3}q^{3}+\mc O(\eta-1)\,.
\ea
The second piece in (\ref{C.30}) is treated similarly. First, we expose the singularity for $\eta\to 1$ by writing 
\ba
\I_{2}^{\rm D3}(\eta^{-1}q^{-\frac{1}{2}}; q^{\frac{3}{2}})  = -\frac{q^{3}}{2\eta^{3}(1-\eta^{3})(1-\eta^{4}q^{2})}\frac{(\eta q^{2})^{2}_{\infty}}{(\frac{q^{2}}{\eta^{2}}; \eta q^{2})_{\infty}(\eta^{5}q^{4}; \eta q^{2})_{\infty}}
\bigg[-3+\frac{2}{\eta^{2}q}R_{1}\bigg(\frac{1}{\eta^{2}q}; \eta q^{2}\bigg)\bigg]\, .
\ea
The combination of q-Pochhammer functions can be expanded around $\eta=1$ using 
\ba
\log& \frac{(\eta q^{2})^{2}_{\infty}}{(\frac{q^{2}}{\eta^{2}}; \eta q^{2})_{\infty}(\eta^{5}q^{4}; \eta q^{2})_{\infty}} \lp
= \sum_{n=0}^{\infty}\bigg[
\log\frac{1-q^{2+2n}}{1-q^{4+2n}}+\bigg(-\frac{(n+4)q^{2+2n}}{1-q^{2+2n}}+\frac{(n+5)q^{4+2n}}{1-q^{4+2n}}\bigg)(\eta-1)+\mc O((\eta-1)^{2})\bigg] \lp
= \log(1-q^{2})-\frac{4q^{2}}{1-q^{2}}(\eta-1)+\mc O((\eta-1)^{2}).
\ea
and therefore
\ba
 \frac{(\eta q^{2})^{2}_{\infty}}{(\frac{q^{2}}{\eta^{2}}; \eta q^{2})_{\infty}(\eta^{5}q^{4}; \eta q^{2})_{\infty}} &= 
  1-q^{2}-4q^{2}(\eta-1)+\mc O((\eta-1)^{2}).
\ea
Hence, 
\ba
\eta^{-2N} & \I_{2}^{\rm D3}(\eta^{-1}q^{-\frac{1}{2}}; q^{\frac{3}{2}})  = -\frac{q^{3}}{3(\eta-1)}\lp
+\frac{1}{3}q\,\bigg[(2+3N)\,q^{2}+q^{3}R_{1}^{(0,1)}(q^{-1}; q^{2})-2R_{1}^{(1,0)}(q^{-1}; q^{2})\bigg]+\mc O(\eta-1)\, ,
\ea
where we used $R_{1}(q^{-1}; q^{2}) = q/2$ and $R_{1}^{(p,q)}$ are partial derivatives. This gives the finite expression 
\be
W_{2}(q) = \bigg(\frac{8}{3}+N\bigg)q^{3}+\frac{q^{4}}{3}R_{1}^{(0,1)}(q^{-1}; q^{2})-\frac{2q}{3}R_{1}^{(1,0)}(q^{-1}; q^{2}).
\ee
Using the differential equation (\ref{C.17}) to work out $R_{1}^{(0,1)}$, we get the expression (\ref{6.15}).

\bibliography{BT-Biblio}
\bibliographystyle{JHEP-v2.9}
\end{document}